\documentclass[lettersize,journal]{IEEEtran}
\usepackage{amsmath,amsfonts}
\usepackage{algcompatible}
\usepackage{algorithm}
\usepackage{array}
\usepackage[caption=false,font=normalsize,labelfont=sf,textfont=sf]{subfig}
\usepackage{textcomp}
\usepackage{stfloats}
\usepackage{url}
\usepackage{verbatim}
\usepackage{graphicx}
\usepackage{cite}
\usepackage{lipsum}
\usepackage{amsmath}
\usepackage{cite}
\usepackage{multicol}

\input{Qcircuit}

\begin{document}

\title{Resource Optimized Quantum Squaring Circuit}

\author{Afrin Sultana, Edgard Mu\~{n}oz-Coreas edgard.munoz-coreas@unt.edu
\thanks{This paper was produced by the IEEE Publication Technology Group. They are in Piscataway, NJ.}
\thanks{Manuscript received April 19, 2021; revised August 16, 2021.}}

\markboth{Journal of \LaTeX\ Class Files,~Vol.~14, No.~8, August~2023}%
{Shell \MakeLowercase{\textit{et al.}}: A Sample Article Using IEEEtran.cls for IEEE Journals}


\maketitle

\begin{abstract}
Quantum squaring operation is a useful building block in implementing quantum algorithms such as linear regression, regularized least squares algorithm, order-finding algorithm, quantum search algorithm, Newton Raphson division, Euclidean distance calculation, cryptography, and in finding roots and reciprocals. Quantum circuits could be made fault-tolerant by using error correcting codes and fault-tolerant quantum gates (such as the Clifford + T-gates). However, the T-gate is very costly to implement. Two qubit gates (such as the CNOT-gate) are more prone to noise errors than single qubit gates. Consequently, in order to realize reliable quantum algorithms, the quantum circuits should have a low T-count and CNOT-count.  In this paper, we present a novel quantum integer squaring architecture optimized for T-count, CNOT-count, T-depth, CNOT-depth, and KQ\textsubscript{T} that produces no garbage outputs. To reduce costs, we use a novel approach for arranging the generated partial products that allows us to reduce the number of adders by 50\%. We also use the resource efficient logical-AND gate and uncomputation gate shown in \cite{C.Gidney.et.al-2018} to further save resources. The proposed quantum squaring circuit sees an asymptotic reduction of 66.67\% in T-count, 50\% in T-depth, 29.41\% in CNOT-count, 42.86\% in CNOT-depth, and 25\% in KQ\textsubscript{T} with respect to Thapliyal et al. \cite{Thapliyal.et.al-2014}. With respect to Nagamani et al. \cite{Nagamani.et.al-2013} the design sees an asymptotic reduction of 77.27\% in T-count, 68.75\% in T-depth, 50\% in CNOT-count, 61.90\% in CNOT-depth, and 6.25\% in the KQ\textsubscript{T}.
\end{abstract}


\section{Introduction}
In the areas of numbers theory, encryption, and scientific computing, quantum algorithms for tasks (such as multivariate integration, path integration, the solution of ordinary and partial differential equations, eigenvalue problems, and numerical linear algebra problems) can  achieve up to a super polynomial factor speedup compared to classical counterparts  \cite{Andersson.et.al-2022}\cite{cambiucci2023hypergraphic}\cite{A.Montanaro.et.al-2023}\cite{W.van.dam.et.al-2008}\cite{park2022optimizing}\cite{van2002efficient}\cite{smith2019simulating}\cite{scolnik2023post}\cite{wang2020quantum}. Quantum implementations for mathematical operations, such as fractional exponents, multiplication, squaring, and addition, will be required to implement many algorithms in these fields \cite{Andersson.et.al-2022}\cite{cambiucci2023hypergraphic}\cite{wang2020quantum}\cite{E.Munoz-Coreas.et.al-2019}\cite{munoz2018t}\cite{bhaskar2015quantum}\cite{duan2019quantum}\cite{nielsen2010quantum}\cite{jayashree2017efficient}\cite{haner2018optimizing}\cite{liang2019quantum}. Specifically, the squaring operation is required to implement quantum algorithms such as linear regression, regularized least squares algorithms, order-finding algorithm, quantum search algorithm, Poisson equation solving, and Euclidean distance calculation \cite{wang2020quantum}\cite{jayashree2017efficient}\cite{demaine2021efficient}\cite{chakraborty2023quantum}. Quantum squaring is also a useful building block to implement higher order functions such as Newton Rhapson division, and reciprocal and root calculation \cite{Thapliyal.et.al-2014}\cite{wang2020quantum}\cite{jayashree2017efficient}. 


Quantum circuits are prone to noise errors. To make quantum circuits resistant to noise errors, fault-tolerant quantum gates and quantum error correcting codes can be applied \cite{E.Munoz-Coreas.et.al-2019}\cite{wu2023optimization}\cite{amy2014polynomial}\cite{bravyi2005universal}\cite{vandaele2023optimal}\cite{vale2023decomposition}. Clifford + T-gates are of interest as they can be made fault-tolerant with error correcting codes \cite{C.Gidney.et.al-2018}\cite{E.Munoz-Coreas.et.al-2019}\cite{miller2014mapping}\cite{amy2013meet}\cite{jones2012layered}\cite{paler2017fault}\cite{devitt2013requirements}\cite{bravyi2005universal}\cite{vandaele2023optimal}. However, the trade-off is that the T-gate is expensive to implement \cite{C.Gidney.et.al-2018}\cite{E.Munoz-Coreas.et.al-2019}\cite{amy2014polynomial}\cite{vandaele2023optimal}\cite{paler2017fault}\cite{devitt2013requirements}\cite{Kento.Oonishi.et.al-2022}\cite{banerjee2014squaring}. Hence, T-count, and T-depth are useful metrics for fault-tolerant quantum circuit design \cite{C.Gidney.et.al-2018}\cite{E.Munoz-Coreas.et.al-2019}\cite{amy2014polynomial}\cite{vandaele2023optimal}\cite{gosset2014algorithm}.  
	

Two qubit gates (such as the CNOT-gate) are more prone to noise errors than single qubit gates \cite{park2022optimizing}\cite{smith2019simulating}\cite{Kento.Oonishi.et.al-2022}\cite{Rigetti}.  As a result, measures such as the CNOT-count and CNOT-depth have caught the attention of researchers \cite{park2022optimizing}\cite{wu2023optimization}\cite{vale2023decomposition}\cite{Kento.Oonishi.et.al-2022}\cite{nash2020quantum}.  Coherence errors (such as the T1 and T2 errors) are a function of time \cite{smith2019simulating}\cite{brand2022markovian}.  Thus, minimizing depth (such as CNOT-depth) has become an important design objective\cite{park2022optimizing}\cite{wu2023optimization}.  Due to the limited qubit resources on quantum machines, the number of qubits is an important resource cost metric\cite{cambiucci2023hypergraphic}. 

The design of quantum squaring circuits have been approached by different researchers in the literature \cite{Thapliyal.et.al-2014}\cite{Nagamani.et.al-2013}\cite{jayashree2017efficient}\cite{banerjee2014squaring}\cite{banerjee2016squaring}. Thapliyal et al. designed a squaring circuit by generating partial products using Toffoli gates in \cite{Thapliyal.et.al-2014}. The partial products are added using Peres gate as half adder and Double Peres gate as full adder. This design has higher resource cost and produces garbage output.  Nagamani et al.  made the design in \cite{Thapliyal.et.al-2014} garbage free by applying Bennett’s garbage removal scheme \cite{bennett1973logical} and presented it as the first design titled \textit{Garbage Free Squaring Unit (GFSU)} in \cite{Nagamani.et.al-2013}. The resulting GFSU has higher gate and depth cost. Nagamani et al. presented another garbage optimized squaring circuit titled  \textit{Optimized Squaring Unit (OSU)}.  The trade-off for this \textit{Optimized Squaring Unit (OSU)} is higher gate and depth costs along with excess garbage output. Banerjee et al. proposed two garbage and ancillae-optimized squaring circuit designs in \cite{banerjee2014squaring} and another reversible squaring circuit using zero garbage and reduced ancillary inputs in \cite{banerjee2016squaring} . The squaring circuits proposed by Banerjee et al. in \cite{banerjee2014squaring} and \cite{banerjee2016squaring} employed recursive structure. As the designs are recursive in nature, they incur higher gate and depth costs compared to other existing works such as \cite{Thapliyal.et.al-2014}.  Lastly, a garbageless reversible squaring circuit based on Karatsuba's recursive method was proposed by Thapliyal et al. in \cite{jayashree2017efficient}. This design employed squaring operation, adder blocks and subtractor blocks recursively for computing and decomputing, which results in higher gate cost than existing designs such as \cite{Thapliyal.et.al-2014}, \cite{Nagamani.et.al-2013}, \cite{banerjee2014squaring}, and \cite{banerjee2016squaring}.

To address the shortcomings of the existing works, we propose a garbageless quantum squaring circuit (QSC) that is optimized for T-count, T-depth, CNOT-count, CNOT-depth, and KQ\textsubscript{T} (the product of the number of qubits and T-depth) and has no garbage output. In this work, a novel technique for the arangement of the generated partial products has been employed. This arrangement allows us to reduce the number of adders used in this circuit by 50\%. Moreover, the design uses the logical-AND gate, uncomputation of logical-AND, and multi-bit quantum adder by Gidney et al. \cite{C.Gidney.et.al-2018} exclusively as building blocks. As a result, the gate and depth cost are further reduced. The proposed QSC, when compared to the existing work, is shown to be superior in terms of T-count, T-depth, CNOT-count, CNOT-depth and KQ\textsubscript{T}.
 
The rest of the paper is organized as follows: In Section 2 we briefly introduce the Clifford + T-gate representations of logical-AND and uncomputation of logical-AND, in Section 3 we present the proposed Quantum Squaring Circuit (QSC) the methodology, Cost Analysis is shown in Section 4, Cost Comparison is shown in Section 5, and in Section 6 we conclude the document.

\section{Background}

\subsection {Quantum Gates}
Physical quantum computers are prone to noise errors. Hence fault-tolerant implementations of quantum circuits are being preferred by researchers. The Clifford + T-gate implementation of quantum circuits could be made fault-tolerant with error correcting codes. In the proposed quantum squaring circuit (QSC) the generation of partial products has been done by using logical-AND gate presented in \cite{C.Gidney.et.al-2018}. The sum of the partial products is computed by using the adder presented in  \cite{C.Gidney.et.al-2018}. Again, the uncomputation of logical-AND has been done. The logical-AND and its uncomputation is illustrated in Fig. 2  \cite{C.Gidney.et.al-2018}.

\begin{figure}[htpb]
\centering
\Qcircuit @C=.8em @R=.2em {
&&\lstick{\ket{x}} & \ctrl{1}\ & \qw\ &  \rstick{\ket{x}} &&&&&&& \qw\ & \ctrl{2}\ & \qw\ & \targ\ & \gate{T^\dag}\ & \targ\ & \qw\ & \qw\ & \qw\\
&&\lstick{\ket{y}} & \ctrl{1}\ & \qw\ & \rstick{\ket{y}} &&&=&&&&\qw\ & \qw\ & \ctrl{1}\ & \targ\ & \gate{T^\dag}\ & \targ\ & \qw\ & \qw\ & \qw\\
&&&&\qw\ & \rstick{\ket{xy}}  &&&&&& \lstick{\ket{T}}  & \qw\ & \targ\ & \targ\ & \ctrl{-2}\ & \gate{T}\ & \ctrl{-2}\ & \gate{H}\ & \gate{S}\ & \qw\
}
\caption{The logical-AND gate and its Clifford + T-gate implementation \cite{C.Gidney.et.al-2018}}

\Qcircuit @C=.8em @R=.2em {
&&&\lstick{\ket{x}} & \ctrl{1}\ & \qw\ &  \rstick{\ket{x}} &&&&& \qw\ & \qw\ & \ctrl{1}\ & \qw\\
&&&\lstick{\ket{y}} & \ctrl{1}\ & \qw\ & \rstick{\ket{y}} &&&=&&\qw\ & \qw\ & \gate{Z}\ & \qw\\
&&&\lstick{\ket{xy}} & \qw\ &&&&&&& \gate{H}\ & \meter\ & \control \cw  \cwx
}
\caption{The uncomputation of logical-AND gate and its Clifford gate implementation \cite{C.Gidney.et.al-2018}}
\end{figure}

\subsection {Evaluation of Quantum Circuit Performance} 
The fault-tolerant implementation of T-gate is more costly than the implementation of other Clifford + T-gates. Hence, the T-count is a useful metric to compare quantum circuits. Other performance metrics include T-depth, CNOT-count, CNOT-depth, and KQ measures.

\begin{itemize}
\item
\textit{T-count}: The T-count is the total number of T-gates or Hermitian transpose of T-gates in a circuit. The T-count of logical-AND is shown in Fig. 1. The number of T-count is 4. 

\item
\textit{T-depth}: The T-depth is the number of T-gate layers in the circuit. A layer is composed of quantum operations that can be performed simultaneously. In the logical-AND circuit, the number of T-depth is 2. 

\textit{CNOT-count} and \textit{CNOT-depth} could be defined similarly. The logical-AND shown in Fig. 1 has a CNOT-count of 6 and CNOT-depth of 4. 

\item
\textit{KQ}: The KQ is the product of qubit cost and gate depth such as T-gate and CNOT-gate.
\end{itemize}


\begin{algorithm*}[htpb]
\begin{multicols}{1}
\caption{Quantum Squaring Circuit}
\begin{algorithmic}[1]
\STATE \textbf{Requirements}: $a$ must be an unsigned integer binary value with bit length $n$ where $n>4$.
\STATE \textbf{Input}: $a$
\STATE \textbf{Output}: $P$ {//The squared value $a^{2}$ is obtained in quantum register $P$ of length $2n$. We will use the variable P to represent $a^{2}$.}
\FOR{$ i$ =$1:n-1$}   {//Generation of parital products and the copying of input bits are implemented in lines 4 through 9}
\FOR{$j$=$i:n-1$}
\STATE $a(i-1,j)=a(i-1).a(j)$   {//AND of inputs values $a\textsubscript{i-1}$ and $a\textsubscript{j}$ results in partial product values $a\textsubscript{i-1,j}$}
\ENDFOR
\STATE $a(i,i)=a(i)$  {//Copy input $a\textsubscript{i}$ to an ancillae using a CNOT-gate. The computation result is $a\textsubscript{i,i}$}
\ENDFOR
\STATE $P(0)=a(0)$ {//$P(0)$ is the least significant bit of $a^{2}$}
\STATE $P(1)=0$ {//$P(1)$ is the second least significant bit of $a^{2}$}
\FOR{$i$ =$1:2n-3$}   {//Lines 12 through 71 show how the partial products and input values are rearranged in order to reduce the adder count by half.}
\IF {$i \leq n-1$ \& $i$ is odd}
\STATE $T(0,i-1)=a((i+1)/2,(i+1)/2)$  {//The input values of  $a\textsubscript{(i+1)/2,(i+1)/2}$ are being place at a quantum register location $T\textsubscript{0,i-1}$.}
\STATE  $T(1,i-1)=a(0,i)$  {//The partial product values of $a\textsubscript{0,i}$ are being placed to quantum register location $T\textsubscript{1,i-1}$}
\IF {$i >1$}
\FOR {$j=2:(i+1)/2$}
\STATE  $T(j,i-2j+1)=a(j-1,i-j+1)$  {//The partial product values of $a\textsubscript{j-1,i-j+1}$ are being placed at quantum register location $T\textsubscript{j,i-2j+1}$.}
\ENDFOR
\ENDIF
\ENDIF
\IF {$i \leq n-1$ \& $i$ is even}
\FOR {$j=1:i/2$}
\IF {$j \leq 2$}
\STATE  $T(j-1,i-1)=a(j-1,i-j+1)$   {//The partial product values of $a\textsubscript{j-1,i-j+1}$ are being placed at quantum register location $T\textsubscript{j-1,i-1}$.}
\ELSE
\STATE  $T(j-1,i-2j+3)=a(j-1,i-j+1)$
\ENDIF
\ENDFOR
\STATE  $T(i/2,1)=0$
\ENDIF
\IF {$i > n-1$ \& $i$ is odd}
\STATE $T(0,i-1)=a((i+1)/2,(i+1)/2)$   {//The input values $a\textsubscript{(i+1)/2,(i+1)/2}$ are placed to quantum register location $T\textsubscript{0,i-1}$.}
\STATE  $T(1,i-1)=a(i-n+1,n-1)$  {//The partial product values $a\textsubscript{i-n+1,n-1}$ are placed to quantum register location $T\textsubscript{1,i-1}$}
\IF {$i \neq (2n-3)$}
\FOR {$j=2:(2n-i-1)/2$}
\STATE  $T(j,i-2j+1)=a(i-n+j,n-j)$   {//The partial product values $a\textsubscript{i-n+j,n-j}$ are placed to quantum register location $T\textsubscript{j,i-2j+1}$.}
\ENDFOR
\ENDIF
\ENDIF
\IF {$i>n-1$ \& $i$ is even}
\FOR {$j=1:(2n-i-2)/2$}
\IF {$j \leq 2$}
\STATE  $T(j-1,i-1)=a(i-n+j,n-j)$   {//Apply the partial product values $a\textsubscript{i-n+j,n-j}$ to quantum register location $T\textsubscript{j-1,i-1}$.}
\ELSE
\STATE  $T(j-1,i-2j+3)=a(i-n+j,n-j)$
\ENDIF
\ENDFOR
\IF{n is odd}
\IF {$i \neq 2n-4$}
\STATE  $T((2n-i-2)/2,2(i-n)+3)=0$   {//Put an ancillae to quantum register location $T\textsubscript{(2n-i-2)/2,2(i-n)+3}$.}
\STATE  $T((2n-i)/2,2(i-n)+1)=0$
\ELSIF{$i = 2n-4$}
\STATE  $T((2n-i-2)/2,i-1)=0$
\STATE  $T((2n-i)/2,i-3)=0$
\ENDIF
\ENDIF
\IF{n is even}
\IF {$i \neq 2n-4$ \& $i \neq n$}
\STATE  $T((2n-i-2)/2,2(i-n)+3)=0$   {//Put an ancillae to quantum register location $T\textsubscript{(2n-i-2)/2,2(i-n)+3}$.}
\STATE  $T((2n-i)/2,2(i-n)+1)=0$
\ELSIF {$i = n$}
\STATE  $T((2n-i-2)/2,3)=0$
\STATE  $T((2n-i)/2,1)=0$
\ELSIF{$i = 2n-4$}
\STATE  $T((2n-i-2)/2,i-1)=0$
\STATE  $T((2n-i)/2,i-3)=0$
\ENDIF
\ENDIF

\ENDIF

\ENDFOR

\IF{n is odd}   {//Ancillae are loaded into quantum register $|T\rangle$. Ancillae are used to ensure properly sized operands are presented to the quantum adders.  Lines 72 through 85 show the ancillae loading procedure.}
\FOR {$i=1:(n-3)/2$}
\FOR {$j=1:2i$}
\STATE  $T(i+1,2n-3-4i+j)=0$
\ENDFOR
\ENDFOR
\ENDIF

\IF{n is even}  
\FOR {$i=1:(n-2)/2$}
\FOR {$j=1:2i$}
\STATE  $T(i+1,2n-3-4i+j)=0$
\ENDFOR
\ENDFOR
\ENDIF

\algstore{testcont}
\end{algorithmic}
\end{multicols}
\end{algorithm*}

\begin{algorithm}[htpb]
\begin{algorithmic}[1]
\algrestore{testcont}

\STATE  $V(0,1:2n-2)=T(0,1:2n-3)+T(1,1:2n-3)$   {//The first addition accepts operands are of length $2n-2$.  Lines 87 through 112 show the remaining additions.  The generation of the squared result P is also shown.}

\IF{n is odd}   
\FOR {$i=1:(n-3)/2$}
\FOR {$j=1:(2n-2-2i)$}
\STATE  $V(i,j-1)=T(i+1,j-1)+V(i-1,j-1)$
\ENDFOR
\IF{$i \neq (n-3)/2$}
\STATE  $P(2i+2)=V(i,0)$ {//Put the first two sum bits to P}
\STATE  $P(2i+3)=V(i,1)$
\ELSE
\STATE  $P(2i+2:2i+n+2)=V(i,0:n)${//Put the last  $n+1$ sum bits to P}	
\ENDIF
\ENDFOR
\ENDIF

\IF{n is even}    
\FOR {$i=1:(n-2)/2$}
\FOR {$j=1:(2n-2-2i)$}
\STATE  $V(i,j-1)=T(i+1,j-1)+V(i-1,j-1)$
\ENDFOR
\IF{$i \neq (n-2)/2$}
\STATE  $P(2i+2)=V(i,0)$  {//Apply the first two sum bits to P}
\STATE  $P(2i+3)=V(i,1)$
\ELSE
\STATE  $P(2i+2:2i+n+1)=V(i,0:n-1)$   {//Put the last n sum bits to P}	
\ENDIF
\ENDFOR
\ENDIF
\STATE $Return$  $P(0:2n-1)$


\end{algorithmic}
\end{algorithm}

\begin{algorithm*}[htpb]
\caption{Uncomputation Algorithm}
\begin{algorithmic}[1]

\FOR {$i=1:2:(2n-3)$}
\STATE  $T(0,i-1)=a((i+1)/2) \oplus T(0,i-1)$   {//Put  a((i+1)/2) and T(0,i-1) to a CNOT gate. a((i+1)/2) is unchanged, T(0,i-1) becomes zero.}	
\ENDFOR

\FOR {$i=3:(2n-3)$}
\IF{$i \leq (n-1)$}
\IF{i is odd}
\STATE  $T(2,i-3)\stackrel{AND^{-1}}\longrightarrow T_{ancillae}$   {//Put a(1), a(i-1) and T(2,i-3) to an uncomputation gate. At the end of computation a(1) and a(i-1) are unchanged and T(2,i-3) is restored to its initial ancillae value}	
\IF{$i > 3$}
\FOR {$i_{1}=1:((i+1)/2-2)$}
\STATE  $T(i_{1}+2,i-3-2i_{1}) \stackrel{AND^{-1}}\longrightarrow T_{ancillae}$
\ENDFOR
\ENDIF
\ENDIF
\IF{i is even}
\STATE  $T(0,i-1) \stackrel{AND^{-1}}\longrightarrow T_{ancillae}$   {//Put a(0), a(i-1) and T(0,i-1) to an uncomputation gate. At the end of computation a(0) and a(i-1) are unchanged and T(0,i-1) is restored to its initial ancillae value}
\IF{$i > 4$}
\FOR {$i_{2}=1:(i/2-2)$}
\STATE  $T(i_{2}+1,i-1-2i_{2}) \stackrel{AND^{-1}}\longrightarrow T_{ancillae}$
\ENDFOR
\ENDIF
\ENDIF
\ENDIF

\IF{$i > (n-1)$}
\IF{i is odd}
\IF{$i \neq 2n-3$}
\FOR {$i_{3}=1:((2n-i-3)/2)$}
\STATE  $T(i_{3}+1,i-1-2i_{3}) \stackrel{AND^{-1}}\longrightarrow T_{ancillae}${//Put $a(i-n+i_{3}+1)$, $a(n-i_{3}-1)$ and $T(i_{3}+1,i-1-2i_{3})$ to an uncomputation gate. At the end of computation $a(i-n+i_{3}+1)$ and $a(n-i_{3}-1)$ are unchanged and $T(i_{3}+1,i-1-2i_{3})$ is restored to its initial ancillae value}
\ENDFOR
\ENDIF
\ENDIF
\IF{i is even}
\STATE  $T(0,i-1) \stackrel{AND^{-1}}\longrightarrow T_{ancillae}$   {//Put a(i-n+1), a(n-1) and T(0,i-1) to an uncomputation gate. At the end of computation a(i-n+1) and a(n-1) are unchanged and T(0,i-1) is restored to its initial ancillae value}
\IF{$i \neq 2n-4 \& i \neq 2n-6$}
\FOR {$i_{4}=2:((2n-i-4)/2)$}
\STATE  $T(i_{4}+1,i-2i_{4}+1)) \stackrel{AND^{-1}}\longrightarrow T_{ancillae}$
\ENDFOR
\ENDIF
\ENDIF
\ENDIF

\ENDFOR

\end{algorithmic}
\end{algorithm*}



\section{Proposed QSC Design}

In this section we show the design methodology of our proposed Quantum Squaring Circuit (QSC) with a lower T-count, T-depth, CNOT-count, CNOT-depth, KQ\textsubscript{T} compared to \cite{Thapliyal.et.al-2014} and \cite{Nagamani.et.al-2013}. To reduce costs, we use a novel approach for arranging the generated partial products which allows us to reduce the number of adders by 50\%. For further reduction, we use the resource efficient logical-AND gate and uncomputation gate shown in \cite{C.Gidney.et.al-2018}. Hence, the gate and depth costs are reduced.  

Consider the squaring of $n$-bit number $a$ stored in quantum register $|A\rangle$. Further, consider a quantum output register $|P\rangle$ of size $2n$ that is initialized to 0.  At the end of computation, quantum registers $|P\rangle$ would have the squared value of $a$. 

The proposed methodology is generic and can design a quantum integer squaring circuit of arbitrary size. The steps involved in the proposed methodology are presented for the squaring of an $n$-bit number $a$. The quantum squaring circuit algorithm is illustrated in algorithm 1. is An illustrative example of the generation of a quantum integer squaring circuit that can perform the squaring of a 6-bit number $a = a\textsubscript{5} a\textsubscript{4} a\textsubscript{3} a\textsubscript{2} a\textsubscript{1} a\textsubscript{0}$ is shown in the figures accompanying the steps below:

\subsection{Steps of QSC Design Methodology}\label{AA}
There are a total of eight steps of the design methodology of our proposed $n$-bit QSC. Steps 1, 2, 3, 4, 7 and 8 are the same for both odd and even values of input $n$. Step 5 is applicable for odd values of input $n$ whereas steps 6 is applicable for even values of input $n$. The uncomputation part of both odd and even values of input $n$ are described in step 8. The steps along with the figures are given below:

\begin{figure}[htpb]
\centering
\Qcircuit @C=0.95em @R=0.8em{
&&&\lstick{\ket{a\textsubscript{0}}} & \ctrl{1} & \ctrl{3} & \ctrl{5} &\ctrl{7} & \ctrl{9} &\qw\ & \qw\ & \qw\ & \qw\ & \qw\ & \rstick{\ket{a\textsubscript{0} \longrightarrow P\textsubscript{0}}} \qw\\
&&&\lstick{\ket{a\textsubscript{1}}} &  \ctrl{1}\ & \qw\ & \qw\ & \qw\ &  \qw\  & \ctrl{10} & \ctrl{2}\  & \ctrl{4} & \ctrl{6} & \ctrl{8}& \qw \\
&&&&& \qw\ & \qw\ & \qw\ & \qw\ &\qw\ & \qw\ & \qw\ & \qw\ & \qw\ &  \rstick{\ket{a\textsubscript{0}a\textsubscript{1}}} \qw\\
&&&\lstick{\ket{a\textsubscript{2}}} & \qw\ & \ctrl{1}\ & \qw\  & \qw\ &\qw\ & \qw\ & \ctrl{9}\ & \qw\ & \qw\ & \qw\ & \qw\\
&&&&&& \qw\ & \qw\ & \qw\ & \qw\ & \qw\ & \qw\  & \qw\ & \qw\ &  \rstick{\ket{a\textsubscript{0}a\textsubscript{2}}} \qw\\
&&&\lstick{\ket{a\textsubscript{3}}} & \qw\ & \qw\ & \ctrl{1}\ & \qw\ & \qw\ & \qw\ & \qw\ &  \ctrl{8}\ & \qw\ & \qw\ & \qw\\
&&&&&&& \qw\ & \qw\ & \qw\ & \qw\ & \qw\ & \qw\  & \qw\ & \rstick{\ket{a\textsubscript{0}a\textsubscript{3}}} \qw\\
&&&\lstick{\ket{a\textsubscript{4}}}  & \qw\ & \qw\ & \qw\ & \ctrl{1}\ & \qw\ & \qw\ & \qw\ & \qw\ &  \ctrl{7}\ & \qw\ &\rstick{\ket{a\textsubscript{0}a\textsubscript{1}}} \qw\\
&&&&&&&& \qw\ & \qw\ & \qw\   & \qw\ & \qw\ & \qw\ & \rstick{\ket{a\textsubscript{0}a\textsubscript{4}}} \qw\\
&&&\lstick{\ket{a\textsubscript{5}}} &  \qw\ & \qw\ & \qw\ & \qw\ & \ctrl{1}\  & \qw\ & \qw\ & \qw\ & \qw\ & \ctrl{6}\ &  \qw\\
&&&&&&&&& \qw\ & \qw\ & \qw\ & \qw\ & \qw\  & \rstick{\ket{a\textsubscript{0}a\textsubscript{5}}} \qw\\
&&&\lstick{\ket{0}} &  \qw\  & \qw\ & \qw\ &  \qw\ & \qw\ &  \targ\ & \qw\ & \qw\ & \qw\ & \qw\ &  \rstick{\ket{a\textsubscript{1}a\textsubscript{1}}} \qw\\
&&&&&&&&&&& \qw\ & \qw\  & \qw\ & \rstick{\ket{a\textsubscript{1}a\textsubscript{2}}} \qw\\
&&&&&&&&&&&& \qw\  &  \qw\ & \rstick{\ket{a\textsubscript{1}a\textsubscript{3}}} \qw\\
&&&&&&&&&&&&&\qw\ & \rstick{\ket{a\textsubscript{1}a\textsubscript{4} }} \qw\\
&&&&&&&&&&&&&&  \rstick{\ket{a\textsubscript{1}a\textsubscript{5}}} \qw
 }
\caption{Step 1: Generation of Partial Products from logical-AND gate for 6-bit input number $a$.  Here, the value $|a\textsubscript{0}\rangle$ is the least significant bit $|P\textsubscript{0}\rangle$ of the squared value of the input $a$.}
\label{fig 3}
\end{figure}
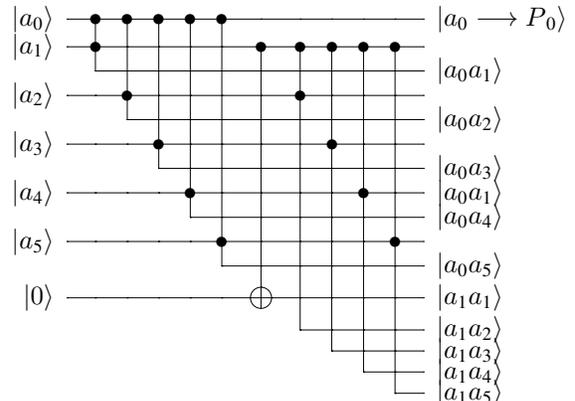
\begin{itemize}
\item Step 1:\\
In this step the input bits of the number $a$ are input to the logical-ANDs. At the end of this step, partial products such as $|a\textsubscript{0}a\textsubscript{1}\rangle$, $|a\textsubscript{0}a\textsubscript{2}\rangle$, $|a\textsubscript{0}a\textsubscript{3}\rangle$... are generated using logical-ANDs. Also input bits are copied and later arranged to feed to adders. The least significant bit $|P\textsubscript{0}\rangle$ is generated in this step. The value $|a\textsubscript{0}\rangle$ is the least significant bit $|P\textsubscript{0}\rangle$ of the squared value of the input $a$.  Here, lines 4 through 11 of algorithm 1 are implemented.\\
\parindent=15.0pt 
For i=1:1:n-1\newline
\indent For j=i:1:n-1\newline
Put the input values $|a\textsubscript{i-1}\rangle$ and $|a\textsubscript{j}\rangle$ at logical-AND gate. The result of the computation is $|a\textsubscript{i-1,j}\rangle$.\newline   
\indent If (i=j), copy input value $|a\textsubscript{i}\rangle$ to an ancillae using a CNOT-gate. The output of the CNOT-gate is the value $|a\textsubscript{i,i}\rangle$.\newline
\end{itemize}
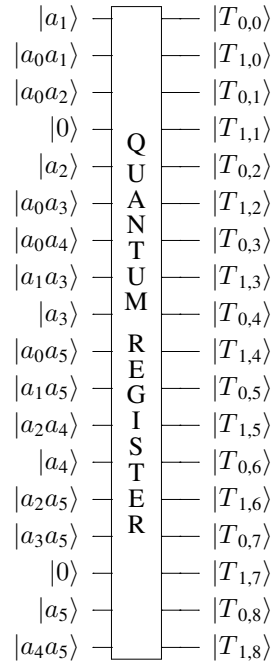
\begin{figure}[htpb]
\centering
\Qcircuit @C=0.7em @R=.5em {
&&&&&&&&\lstick{\ket{a\textsubscript{1}}} & \multigate{17}{\mathcal{\shortstack{Q\\U\\A\\N\\T\\U\\M\\\\\\R\\E\\G\\I\\S\\T\\E\\R}}} & \qw\ & \rstick{\ket{T\textsubscript{0,0}}} \qw\\
&&&&&&&&\lstick{\ket{a\textsubscript{0}a\textsubscript{1}}}   & \ghost{\mathcal{\shortstack{Q\\U\\A\\N\\T\\U\\M\\\\\\R\\E\\G\\I\\S\\T\\E\\R}}}& \qw\ & \rstick{\ket{T\textsubscript{1,0}}} \qw\\
&&&&&&&&\lstick{\ket{a\textsubscript{0}a\textsubscript{2}}}   & \ghost{\mathcal{\shortstack{Q\\U\\A\\N\\T\\U\\M\\\\\\R\\E\\G\\I\\S\\T\\E\\R}}}& \qw\ & \rstick{\ket{T\textsubscript{0,1}}} \qw\\
&&&&&&&&\lstick{\ket{0}}  & \ghost{\mathcal{\shortstack{Q\\U\\A\\N\\T\\U\\M\\\\\\R\\E\\G\\I\\S\\T\\E\\R}}} & \qw\ & \rstick{\ket{T\textsubscript{1,1}}} \qw\\
&&&&&&&&\lstick{\ket{a\textsubscript{2}}}  & \ghost{\mathcal{\shortstack{Q\\U\\A\\N\\T\\U\\M\\\\\\R\\E\\G\\I\\S\\T\\E\\R}}} & \qw\ & \rstick{\ket{T\textsubscript{0,2}}} \qw\\
&&&&&&&&\lstick{\ket{a\textsubscript{0}a\textsubscript{3}}} & \ghost{\mathcal{\shortstack{Q\\U\\A\\N\\T\\U\\M\\\\\\R\\E\\G\\I\\S\\T\\E\\R}}}  & \qw\ & \rstick{\ket{T\textsubscript{1,2}}} \qw\\
&&&&&&&&\lstick{\ket{a\textsubscript{0}a\textsubscript{4}}}  & \ghost{\mathcal{\shortstack{Q\\U\\A\\N\\T\\U\\M\\\\\\R\\E\\G\\I\\S\\T\\E\\R}}} & \qw\ & \rstick{\ket{T\textsubscript{0,3}}} \qw\\
&&&&&&&&\lstick{\ket{a\textsubscript{1}a\textsubscript{3}}}  & \ghost{\mathcal{\shortstack{Q\\U\\A\\N\\T\\U\\M\\\\\\R\\E\\G\\I\\S\\T\\E\\R}}} & \qw\ & \rstick{\ket{T\textsubscript{1,3}}} \qw\\
&&&&&&&&\lstick{\ket{a\textsubscript{3}}}  & \ghost{\mathcal{\shortstack{Q\\U\\A\\N\\T\\U\\M\\\\\\R\\E\\G\\I\\S\\T\\E\\R}}} & \qw\ & \rstick{\ket{T\textsubscript{0,4}}} \qw\\
&&&&&&&&\lstick{\ket{a\textsubscript{0}a\textsubscript{5}}}  & \ghost{\mathcal{\shortstack{Q\\U\\A\\N\\T\\U\\M\\\\\\R\\E\\G\\I\\S\\T\\E\\R}}} & \qw\ & \rstick{\ket{T\textsubscript{1,4}}} \qw\\
&&&&&&&&\lstick{\ket{a\textsubscript{1}a\textsubscript{5}}}  & \ghost{\mathcal{\shortstack{Q\\U\\A\\N\\T\\U\\M\\\\\\R\\E\\G\\I\\S\\T\\E\\R}}} & \qw\ & \rstick{\ket{T\textsubscript{0,5}}} \qw\\
&&&&&&&&\lstick{\ket{a\textsubscript{2}a\textsubscript{4}}}  & \ghost{\mathcal{\shortstack{Q\\U\\A\\N\\T\\U\\M\\\\\\R\\E\\G\\I\\S\\T\\E\\R}}} & \qw\ & \rstick{\ket{T\textsubscript{1,5}}} \qw\\
&&&&&&&&\lstick{\ket{a\textsubscript{4}}} & \ghost{\mathcal{\shortstack{Q\\U\\A\\N\\T\\U\\M\\\\\\R\\E\\G\\I\\S\\T\\E\\R}}}  & \qw\ & \rstick{\ket{T\textsubscript{0,6}}} \qw\\
&&&&&&&&\lstick{\ket{a\textsubscript{2}a\textsubscript{5}}}  & \ghost{\mathcal{\shortstack{Q\\U\\A\\N\\T\\U\\M\\\\\\R\\E\\G\\I\\S\\T\\E\\R}}} & \qw\ & \rstick{\ket{T\textsubscript{1,6}}} \qw\\
&&&&&&&&\lstick{\ket{a\textsubscript{3}a\textsubscript{5}}}  & \ghost{\mathcal{\shortstack{Q\\U\\A\\N\\T\\U\\M\\\\\\R\\E\\G\\I\\S\\T\\E\\R}}} & \qw\ & \rstick{\ket{T\textsubscript{0,7}}} \qw\\
&&&&&&&&\lstick{\ket{0}}  & \ghost{\mathcal{\shortstack{Q\\U\\A\\N\\T\\U\\M\\\\\\R\\E\\G\\I\\S\\T\\E\\R}}} & \qw\ & \rstick{\ket{T\textsubscript{1,7}}} \qw\\
&&&&&&&&\lstick{\ket{a\textsubscript{5}}}  & \ghost{\mathcal{\shortstack{Q\\U\\A\\N\\T\\U\\M\\\\\\R\\E\\G\\I\\S\\T\\E\\R}}} & \qw\ & \rstick{\ket{T\textsubscript{0,8}}} \qw\\
&&&&&&&&\lstick{\ket{a\textsubscript{4}a\textsubscript{5}}}   & \ghost{\mathcal{\shortstack{Q\\U\\A\\N\\T\\U\\M\\\\\\R\\E\\G\\I\\S\\T\\E\\R}}}& \qw\ & \rstick{\ket{T\textsubscript{1,8}}} \qw\
}
\caption{Step 2: The arranged partial products along with ancillae is placed at  a quantum register $|T\rangle$ for  6-bit input number $a$ is shown above. For example, the generated partial product $|a\textsubscript{0}a\textsubscript{1}\rangle$ is placed at quantum register location $|T\textsubscript{1,0}\rangle$, $|$a\textsubscript{0}a\textsubscript{1}$\rightarrow$T\textsubscript{1,0}$\rangle$.}
\label{fig 4}
\end{figure}
\begin{itemize}
\item Step 2:\\
In this step the generated partial products in step 1 are arranged specifically and placed at temporary quantum register $|T\rangle$. This arrangement of generated partial products is done to reduce the number of the adders to half of the number used in existing designs \cite{Thapliyal.et.al-2014} and \cite{Nagamani.et.al-2013}. Values of $|T\textsubscript{0,j}\rangle$ and $|T\textsubscript{1,j}\rangle$ where $j=0,1,2...2n-2$ from the array are feed to quantum adder in step 4, and the remaining partial products from the array are feed to the quantum adder in step 5 for $n$ odd and step 6 for $n$ even. In this step lines 12 through 71 of algorithm 1 is implemented.\\

For i=1:2n-3 \newline
\begin{itemize}
\item Sub-Step 1:\\
\parindent=15.0pt
\indent If $i\leq n-1$ and i is odd then put the input bit $|a\textsubscript{(i+1)/2,(i+1)/2}\rangle$ to quantum register location $|T\textsubscript{0,i-1}\rangle$ and the partial product of logical-AND value $|a\textsubscript{0,i}\rangle$ to quantam register location $|T\textsubscript{1,i-1}\rangle$ respectively.\newline
\begin{itemize}
\item Sub-sub-step 1: 
If $i>1$ and odd then, \\
For j=2:(i+1)/2\\
Put the partial product of logical-AND value $|a\textsubscript{j-1,(i-j+1)}\rangle$ to quantum register location $|T\textsubscript{j,i-2j+1)}\rangle$. Here line 18 of algorithm 1 is implemented.\newline
\end{itemize}
\item Sub-Step 2:
If $i\leq n-1$ and i is even then, (Here lines 22 through 31 of algortihm 1 are implemented).\\
For j=1:i/2\\
\parindent=15.0pt
\indent If $j\leq2$ then apply the partial product of the logical-AND value $|a\textsubscript{j-1,i-j+1}\rangle$ at quantum register location $|T\textsubscript{j-1,i-1}\rangle$.\\
\parindent=15.0pt
\indent If $j>2$ then apply the partial product of the logical-AND value $|a\textsubscript{j-1,i-j+1}\rangle$ at quantum register location $|T\textsubscript{j-1,i-2j+3}\rangle$.\\
\noindent Put an ancillae to $|T\textsubscript{i/2,1}\rangle$.\\ 
\item Sub-Step 3:
 If $i>n-1$ and i is odd then,\\
 Apply input bit $|a\textsubscript{(i+1)/2,(i+1)/2}\rangle$ to quantum register location $|T\textsubscript{0,i-1}\rangle$ and the partial product of logical-AND value
 $|a\textsubscript{i-n+1,n-1}\rangle$ to quantum register location $|T\textsubscript{1,i-1}\rangle$.\newline
\begin{itemize}
\item Sub-sub-step 1: 
If $i\neq 2n-3$ then\\
For j=2:(2n-i-1)/2\\
Put the partial product of logical-AND value $|a\textsubscript{i-n+j,n-j}\rangle$ to quantum register location $|T\textsubscript{j,i-2j+1)}\rangle$. Here line 37 of algorithm 1 is implemented.\newline
\end{itemize}
\item Sub-Step 4:
If $i>n-1$ and i is even then,\\
For j=1:(2n-i-2)/2\\
\parindent=15pt
\indent If $j\leq2$ then put the partial product of the logical-AND value $|a\textsubscript{i-n+j,n-j}\rangle$ at quantum register location $|T\textsubscript{j-1,i-1}\rangle$.\\
\indent If $j>2$ then put the partial product of the logical-AND value $|a\textsubscript{i-n+j,n-j}\rangle$ at quantum register location $|T\textsubscript{j-1,i-2j+3}\rangle$.\\

\begin{itemize}
\item Sub-Sub-Step 1:\\
Here lines 49 through 57 of algorithm 1 are implemented\\  
If n is odd then\\
\parindent=15.0pt 
\indent If $i\neq2n-4$ then \\
Assign two ancillae to quantum register location $|T\textsubscript{(2n-i-2)/2,2(i-n)+3}\rangle$ and $|T\textsubscript{(2n-i)/2,2(i-n)+1}\rangle$.\\
\indent If $i=2n-4$ then \\
Assign two ancillae to quantum register location $|T\textsubscript{(2n-i-2)/2,i-1}\rangle$ and $|T\textsubscript{(2n-i)/2,i-3}\rangle$.\\
\item Sub-Sub-Step 2:\\
Here lines 58 through 71 of algorithm 1 are implemented\\ 
If n is even then\\
\indent  If $i\neq2n-4 \& i\neq n$ then \\
Assign two ancillae to quantum register location $|T\textsubscript{(2n-i-2)/2,2(i-n)+3}\rangle$ and $|T\textsubscript{(2n-i)/2,2(i-n)+1}\rangle$.\\
\indent  If $i=2n-4$ then \\
Assign two ancillae to quantum register location $|T\textsubscript{(2n-i-2)/2,i-1}\rangle$ and $|T\textsubscript{(2n-i)/2,i-3}\rangle$.\\
\indent If $i=n$ then \\
Assign two ancillae to quantum register location $|T\textsubscript{(2n-i-2)/2,3}\rangle$ and $|T\textsubscript{(2n-i)/2,1}\rangle$.\\

\end{itemize}     

\end{itemize}
\end{itemize}

\begin{figure}[htpb]
\centering
\Qcircuit @C=0.7em @R=0.5em {
&&&&&&&&\lstick{\ket{a\textsubscript{1}a\textsubscript{2}}} & \multigate{13}{\mathcal{\shortstack{Q\\U\\A\\N\\T\\U\\M\\\\\\R\\E\\G\\I\\S\\T\\E\\R}}}  & \qw\ & \rstick{\ket{T\textsubscript{2,0}}} \qw\\
&&&&&&&&\lstick{\ket{0}}  & \ghost{\mathcal{\shortstack{Q\\U\\A\\N\\T\\U\\M\\\\\\R\\E\\G\\I\\S\\T\\E\\R}}} & \qw\ & \rstick{\ket{T\textsubscript{2,1}}} \qw\\
&&&&&&&&\lstick{\ket{a\textsubscript{1}a\textsubscript{4}}}  & \ghost{\mathcal{\shortstack{Q\\U\\A\\N\\T\\U\\M\\\\\\R\\E\\G\\I\\S\\T\\E\\R}}} & \qw\ & \rstick{\ket{T\textsubscript{2,2}}} \qw\\
&&&&&&&&\lstick{\ket{0}}  & \ghost{\mathcal{\shortstack{Q\\U\\A\\N\\T\\U\\M\\\\\\R\\E\\G\\I\\S\\T\\E\\R}}} & \qw\ & \rstick{\ket{T\textsubscript{2,3}}} \qw\\
&&&&&&&&\lstick{\ket{a\textsubscript{3}a\textsubscript{4}}}  & \ghost{\mathcal{\shortstack{Q\\U\\A\\N\\T\\U\\M\\\\\\R\\E\\G\\I\\S\\T\\E\\R}}} & \qw\ & \rstick{\ket{T\textsubscript{2,4}}} \qw\\
&&&&&&&&\lstick{\ket{0}} & \ghost{\mathcal{\shortstack{Q\\U\\A\\N\\T\\U\\M\\\\\\R\\E\\G\\I\\S\\T\\E\\R}}}  & \qw\ & \rstick{\ket{T\textsubscript{2,5}}} \qw\\
&&&&&&&&\lstick{\ket{0}} & \ghost{\mathcal{\shortstack{Q\\U\\A\\N\\T\\U\\M\\\\\\R\\E\\G\\I\\S\\T\\E\\R}}}  & \qw\ & \rstick{\ket{T\textsubscript{2,6}}} \qw\\
&&&&&&&&\lstick{\ket{0}}  & \ghost{\mathcal{\shortstack{Q\\U\\A\\N\\T\\U\\M\\\\\\R\\E\\G\\I\\S\\T\\E\\R}}} & \qw\ & \rstick{\ket{T\textsubscript{2,7}}} \qw\\
&&&&&&&&\lstick{\ket{a\textsubscript{2}a\textsubscript{3}}}  & \ghost{\mathcal{\shortstack{Q\\U\\A\\N\\T\\U\\M\\\\\\R\\E\\G\\I\\S\\T\\E\\R}}} & \qw\ & \rstick{\ket{T\textsubscript{3,0}}} \qw\\
&&&&&&&&\lstick{\ket{0}} & \ghost{\mathcal{\shortstack{Q\\U\\A\\N\\T\\U\\M\\\\\\R\\E\\G\\I\\S\\T\\E\\R}}}  & \qw\ & \rstick{\ket{T\textsubscript{3,1}}} \qw\\
&&&&&&&&\lstick{\ket{0}} & \ghost{\mathcal{\shortstack{Q\\U\\A\\N\\T\\U\\M\\\\\\R\\E\\G\\I\\S\\T\\E\\R}}}  & \qw\ & \rstick{\ket{T\textsubscript{3,2}}} \qw\\
&&&&&&&&\lstick{\ket{0}} & \ghost{\mathcal{\shortstack{Q\\U\\A\\N\\T\\U\\M\\\\\\R\\E\\G\\I\\S\\T\\E\\R}}}  & \qw\ & \rstick{\ket{T\textsubscript{3,3}}} \qw\\
&&&&&&&&\lstick{\ket{0}} & \ghost{\mathcal{\shortstack{Q\\U\\A\\N\\T\\U\\M\\\\\\R\\E\\G\\I\\S\\T\\E\\R}}}  & \qw\ & \rstick{\ket{T\textsubscript{3,4}}} \qw\\
&&&&&&&&\lstick{\ket{0}} & \ghost{\mathcal{\shortstack{Q\\U\\A\\N\\T\\U\\M\\\\\\R\\E\\G\\I\\S\\T\\E\\R}}}  & \qw\ & \rstick{\ket{T\textsubscript{3,5}}} \qw\
}
\caption{Step 3: The remaining array of generated partial products from step 2 are zero padded and placed to a quantum register $|T\rangle$ for 6-bit input number $a$. For example,  $|a\textsubscript{1}a\textsubscript{2}\rangle$ value is put at quantum register location value $|T\textsubscript{2,0}\rangle$, $|$a\textsubscript{1}a\textsubscript{2}$\rightarrow$T\textsubscript{2,0}$\rangle$.}
\label{fig 6}
\end{figure}
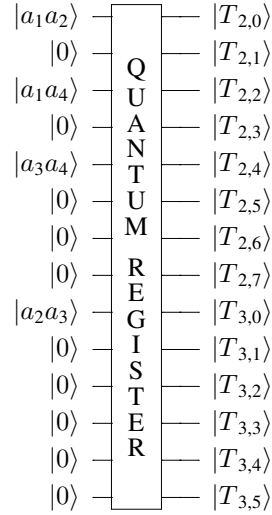

\begin{itemize}	
\item Step 3: \\
In this step zero-padding (loading ancillae) is done to the remaining locations of $|T\rangle$. This is done to ensure that properly sized operands are presented to the quantum adders.  Lines 72 through 85 of algorithm 1 are implemented in this step.
\newline
\begin{itemize}
\item Sub-step 1: Here lines 72-78 are implemented.\\
If n is odd then\\
\parindent=15pt
\indent For i=1:(n-3)/2\newline
\parindent=25.0pt 
\indent For j=1:2i\newline
Assign an ancillae to quantum register location $|T\textsubscript{i+1,2n-4i+j-3}\rangle$.\newline

\item Sub-step 2: Here lines 79-85 are implemented.\\
If n is even the\\
\parindent=15.0pt
\indent For i=1:(n-2)/2\newline
\parindent=25.0pt
\indent For j=1:2i\newline
Assign an ancillae to quantum register location $|T\textsubscript{i+1,2n-4i+j-3}\rangle$.\newline

\end{itemize}



\end{itemize}

These rearranged partial products are then placed at a quantum register $|T\rangle$ for 6-bit number $a$ as shown in Fig. 4 and 5.

\begin{itemize}
\item Step 4:\\
In this step, values $|T\textsubscript{0,0}\rangle$ to $|T\textsubscript{0,2n-2}\rangle$, and values  $|T\textsubscript{1,0}\rangle$ to $|T\textsubscript{1,2n-2}\rangle$ are applied as $2n-3$ bit operands to a $2n-3$ bit quantum adder.  At the end of computation, $2n-2$ bit sum is produced.  The sum bits $|S\textsubscript{0}\rangle$, $|S\textsubscript{1}\rangle$ are bits $|P\textsubscript{2}\rangle$, and $|P\textsubscript{3}\rangle$ of the squared value of input $a$. The remaining $2n-4$ sum output bits are placed at quantum register $|V\rangle$. In this step line 86 of algorithm 1 is implemented.
\end{itemize}

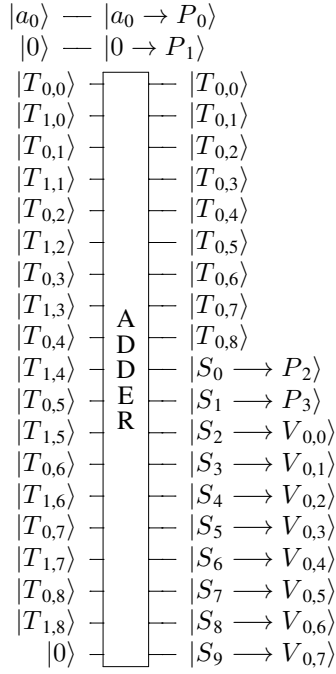
\begin{figure}[htpb]
\centering
\Qcircuit @C=0.5em @R=0.3em {
&&&&&&&&&&&&&&&&&&&&&\lstick{\ket{a\textsubscript{0}}} & \qw\  & \rstick{\ket{a\textsubscript{0} \rightarrow P\textsubscript{0}}} \qw\\
\\
\\
\\
&&&&&&&&&&&&&&&&&&&&&\lstick{\ket{0}} &  \qw\ & \rstick{\ket{0 \rightarrow P\textsubscript{1}}} \qw\\
\\
\\
&&&&&&&&&&&&&&&&&&&&&&&\lstick{\ket{T\textsubscript{0,0}}} & \multigate{18}{\mathcal{\shortstack{A\\D\\D\\E\\R}}}& \qw\ & \rstick{\ket{T\textsubscript{0,0}}} \qw \\
&&&&&&&&&&&&&&&&&&&&&&&\lstick{\ket{T\textsubscript{1,0}}}& \ghost{\mathcal{\shortstack{A\\D\\D\\E\\R}}} & \qw\ &\rstick{\ket{T\textsubscript{0,1}}} \qw\\
&&&&&&&&&&&&&&&&&&&&&&&\lstick{\ket{T\textsubscript{0,1}}}& \ghost{\mathcal{\shortstack{A\\D\\D\\E\\R}}} & \qw\ &\rstick{\ket{T\textsubscript{0,2}}} \qw\\
&&&&&&&&&&&&&&&&&&&&&&&\lstick{\ket{T\textsubscript{1,1}}}& \ghost{\mathcal{\shortstack{A\\D\\D\\E\\R}}} &  \qw\ &\rstick{\ket{T\textsubscript{0,3}}} \qw\\
&&&&&&&&&&&&&&&&&&&&&&&\lstick{\ket{T\textsubscript{0,2}}}& \ghost{\mathcal{\shortstack{A\\D\\D\\E\\R}}}& \qw\ & \rstick{\ket{T\textsubscript{0,4}}} \qw\\ 
&&&&&&&&&&&&&&&&&&&&&&&\lstick{\ket{T\textsubscript{1,2}}}& \ghost{\mathcal{\shortstack{A\\D\\D\\E\\R}}} & \qw\ & \rstick{\ket{T\textsubscript{0,5}}} \qw\\ 
&&&&&&&&&&&&&&&&&&&&&&&\lstick{\ket{T\textsubscript{0,3}}}& \ghost{\mathcal{\shortstack{A\\D\\D\\E\\R}}} &\qw\ & \rstick{\ket{T\textsubscript{0,6}}}\qw\\
&&&&&&&&&&&&&&&&&&&&&&&\lstick{\ket{T\textsubscript{1,3}}}& \ghost{\mathcal{\shortstack{A\\D\\D\\E\\R}}}  & \qw\  &   \rstick{\ket{T\textsubscript{0,7}}}\qw\\
&&&&&&&&&&&&&&&&&&&&&&&\lstick{\ket{T\textsubscript{0,4}}}& \ghost{\mathcal{\shortstack{ A\\D\\D\\E\\R}}}  &\qw\ &  \rstick{\ket{T\textsubscript{0,8}}}\qw\\
&&&&&&&&&&&&&&&&&&&&&&&\lstick{\ket{T\textsubscript{1,4}}}& \ghost{\mathcal{\shortstack{A\\D\\D\\E\\R}}}  & \qw\ &  \rstick{\ket{S\textsubscript{0}\longrightarrow P\textsubscript{2}}}\qw\\
&&&&&&&&&&&&&&&&&&&&&&&\lstick{\ket{T\textsubscript{0,5}}}& \ghost{\mathcal{\shortstack{A\\D\\D\\E\\R}}}  & \qw\ &   \rstick{\ket{S\textsubscript{1}\longrightarrow P\textsubscript{3}}}\qw\\
&&&&&&&&&&&&&&&&&&&&&&&\lstick{\ket{T\textsubscript{1,5}}}& \ghost{\mathcal{\shortstack{A\\D\\D\\E\\R}}}  & \qw\ &   \rstick{\ket{S\textsubscript{2}\longrightarrow V\textsubscript{0,0}}}\qw\\
&&&&&&&&&&&&&&&&&&&&&&&\lstick{\ket{T\textsubscript{0,6}}}& \ghost{\mathcal{\shortstack{A\\D\\D\\E\\R}}}  &\qw\ &   \rstick{\ket{S\textsubscript{3}\longrightarrow V\textsubscript{0,1}}}\qw \\
&&&&&&&&&&&&&&&&&&&&&&&\lstick{\ket{T\textsubscript{1,6}}}& \ghost{\mathcal{\shortstack{A\\D\\D\\E\\R}}}  & \qw\ &   \rstick{\ket{S\textsubscript{4}\longrightarrow V\textsubscript{0,2}}}\qw \\
&&&&&&&&&&&&&&&&&&&&&&&\lstick{\ket{T\textsubscript{0,7}}}& \ghost{\mathcal{\shortstack{A\\D\\D\\E\\R}}}  & \qw\ &   \rstick{\ket{S\textsubscript{5}\longrightarrow V\textsubscript{0,3}}}\qw \\
&&&&&&&&&&&&&&&&&&&&&&&\lstick{\ket{T\textsubscript{1,7}}}& \ghost{\mathcal{\shortstack{A\\D\\D\\E\\R}}}  & \qw\ &   \rstick{\ket{S\textsubscript{6}\longrightarrow V\textsubscript{0,4}}}\qw \\
&&&&&&&&&&&&&&&&&&&&&&&\lstick{\ket{T\textsubscript{0,8}}}& \ghost{\mathcal{\shortstack{A\\D\\D\\E\\R}}}  & \qw\ &   \rstick{\ket{S\textsubscript{7}\longrightarrow V\textsubscript{0,5}}}\qw \\
&&&&&&&&&&&&&&&&&&&&&&&\lstick{\ket{T\textsubscript{1,8}}}& \ghost{\mathcal{\shortstack{A\\D\\D\\E\\R}}}  &  \qw\ &   \rstick{\ket{S\textsubscript{8}\longrightarrow V\textsubscript{0,6}}}\qw \\
&&&&&&&&&&&&&&&&&&&&&&& \lstick{\ket{0}} & \ghost{\mathcal{\shortstack{A\\D\\D\\E\\R}}}  & \qw\ &   \rstick{\ket{S\textsubscript{9}\longrightarrow V\textsubscript{0,7}}}\qw\
}
\caption{Step 4: The arranged partial products of 6-bit input number $a$ obtained from logical-ANDs are input to $2n-3$ bit adder. The sum values $|S\rangle$ are the outputs from the adder. $|P\rangle$ values are the squared qubits of input $a$. The value of $|P\textsubscript{1}\rangle$ of the squared qubits ($a^{2}$) of input $a$ is from ancillae $|0\rangle$. The sum bits  $|S\textsubscript{0}\rangle$, $|S\textsubscript{1}\rangle$ are bits $|P\textsubscript{2}\rangle$ $|P\textsubscript{3}\rangle$ of the squared value $|a\textsubscript{2}\rangle$. Along with the sum bits the input $|T\textsubscript{0,0}\rangle$ to $|T\textsubscript{0,8}\rangle$ are regenerated for 6-bit input number $a$.}
\label{fig 5}
\end{figure}

\begin{itemize}
\item Step 5:\\
Step 5 takes place for $n$ odd.  In this step, $(n-3)/2$ additions are to be performed.  For each addition, two bits of the final squared value of $a$ are generated.  Each adder takes as input partial products stored in register $|T\rangle$ and the remaining sum bits from the last round of computation which is stored in quantum register $|V\rangle$.  The following executes lines 87 through 99 of algorithm 1.\newline
For i=1:(n-3)/2\newline
\parindent=15.0pt 
\indent For j=2n-2-2i
\begin{itemize}
\item Sub-Step 1: 
Apply $|V\textsubscript{i-1,0}\rangle$ to $|V\textsubscript{i-1,j-1}\rangle$  sum bits and $|T\textsubscript{i+1,0}\rangle$ to $|T\textsubscript{i+1,j-1}\rangle$ zero-padded partial products to $2n-2-2i$-bit  adder. On each addition operation $|P\textsubscript{(2i+2)}\rangle$, $|P\textsubscript{(2i+3)}\rangle$ bits of the squared value  $|a\textsuperscript{2}\rangle$ are generated from the first two sum bits of the $2n-1-2i$ sum bits. The generated carry bit is redundant.
\item Sub-Step 2:
Place the remaining $2n-4-2i$ sum bits to quantum register $|V\textsubscript{i,0}\rangle$ to $|V\textsubscript{i,2n-5-2i}\rangle$. Repeat Sub-step 1.\\
\noindent If i=(n-3)/2 then $|P\textsubscript{2i+2}\rangle$ to $|P\textsubscript{2i+2+n}\rangle$ bits of the squared value  $|a\textsuperscript{2}\rangle$ are generated from $|V\textsubscript{i,0}\rangle$ to $|V\textsubscript{i,n}\rangle$. 
\end{itemize}
\end{itemize}

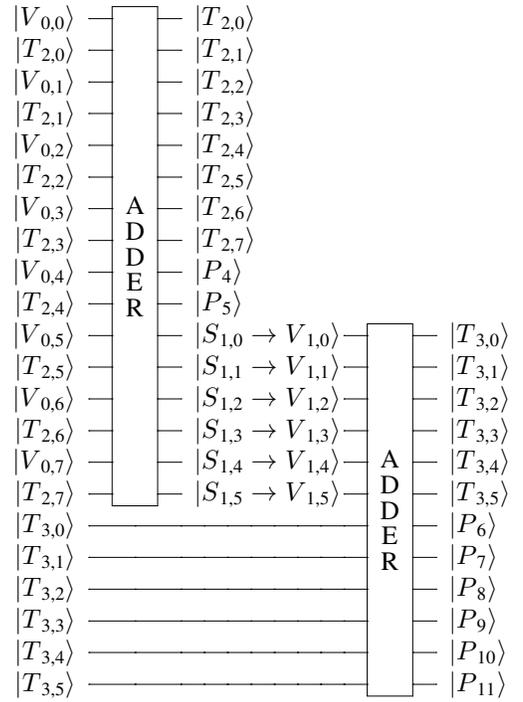
\begin{figure}[htpb]
\centering
\Qcircuit @C=.88em @R=.3em{
&&&&&&&&\lstick{\ket{V\textsubscript{0,0}}} & \multigate{15}{\mathcal{\shortstack{A\\D\\D\\E\\R}}}& \rstick{\ket{T\textsubscript{2,0}}}  \qw\\
&&&&&&&&\lstick{\ket{T\textsubscript{2,0}}}& \ghost{\mathcal{\shortstack{A\\D\\D\\E\\R}}} & \rstick{\ket{T\textsubscript{2,1}}}  \qw\\
&&&&&&&&\lstick{\ket{V\textsubscript{0,1}}}& \ghost{\mathcal{\shortstack{A\\D\\D\\E\\R}}}& \rstick{\ket{T\textsubscript{2,2}}}  \qw\\
&&&&&&&&\lstick{\ket{T\textsubscript{2,1}}}& \ghost{\mathcal{\shortstack{A\\D\\D\\E\\R}}}& \rstick{\ket{T\textsubscript{2,3}}}  \qw \\
&&&&&&&&\lstick{\ket{V\textsubscript{0,2}}}& \ghost{\mathcal{\shortstack{A\\D\\D\\E\\R}}}& \rstick{\ket{T\textsubscript{2,4}}}  \qw \\
&&&&&&&&\lstick{\ket{T\textsubscript{2,2}}}& \ghost{\mathcal{\shortstack{A\\D\\D\\E\\R}}}& \rstick{\ket{T\textsubscript{2,5}}}  \qw \\
&&&&&&&&\lstick{\ket{V\textsubscript{0,3}}}& \ghost{\mathcal{\shortstack{A\\D\\D\\E\\R}}}& \rstick{\ket{T\textsubscript{2,6}}}  \qw\\
&&&&&&&&\lstick{\ket{T\textsubscript{2,3}}} & \ghost{\mathcal{\shortstack{A\\D\\D\\E\\R}}}& \rstick{\ket{T\textsubscript{2,7}}}  \qw\\
&&&&&&&&\lstick{\ket{V\textsubscript{0,4}}}& \ghost{\mathcal{\shortstack{A\\D\\D\\E\\R}}}& \rstick{\ket{P\textsubscript{4}}}  \qw\\
&&&&&&&&\lstick{\ket{T\textsubscript{2,4}}}&\ghost{\mathcal{\shortstack{A\\D\\D\\E\\R}}}& \rstick{\ket{P\textsubscript{5}}} \qw\\
&&&&&&&&\lstick{\ket{V\textsubscript{0,5}}}& \ghost{\mathcal{\shortstack{A\\D\\D\\E\\R}}} & \rstick{\ket{S\textsubscript{1,0} \rightarrow V\textsubscript{1,0}}}  \qw\ &&&&&&&& \multigate{11}{\mathcal{\shortstack{A\\D\\D\\E\\R}}} & \rstick{\ket{T\textsubscript{3,0}}}  \qw\\
&&&&&&&&\lstick{\ket{T\textsubscript{2,5}}}& \ghost{\mathcal{\shortstack{A\\D\\D\\E\\R}}} & \rstick{\ket{S\textsubscript{1,1} \rightarrow V\textsubscript{1,1}}} \qw\ &&&&&&&&\ghost{\mathcal{\shortstack{A\\D\\D\\E\\R}}} & \rstick{\ket{T\textsubscript{3,1}}}  \qw\\
&&&&&&&&\lstick{\ket{V\textsubscript{0,6}}}& \ghost{\mathcal{\shortstack{A\\D\\D\\E\\R}}} & \rstick{\ket{S\textsubscript{1,2} \rightarrow V\textsubscript{1,2}}} \qw\ &&&&&&&&\ghost{\mathcal{\shortstack{A\\D\\D\\E\\R}}}& \rstick{\ket{T\textsubscript{3,2}}}  \qw\\
&&&&&&&&\lstick{\ket{T\textsubscript{2,6}}}& \ghost{\mathcal{\shortstack{A\\D\\D\\E\\R}}} & \rstick{\ket{S\textsubscript{1,3} \rightarrow V\textsubscript{1,3}}} \qw\ &&&&&&&&\ghost{\mathcal{\shortstack{A\\D\\D\\E\\R}}}& \rstick{\ket{T\textsubscript{3,3}}}  \qw\\
&&&&&&&&\lstick{\ket{V\textsubscript{0,7}}}& \ghost{\mathcal{\shortstack{A\\D\\D\\E\\R}}} & \rstick{\ket{S\textsubscript{1,4} \rightarrow V\textsubscript{1,4}}} \qw\ &&&&&&&&\ghost{\mathcal{\shortstack{A\\D\\D\\E\\R}}}& \rstick{\ket{T\textsubscript{3,4}}}  \qw\\
&&&&&&&&\lstick{\ket{T\textsubscript{2,7}}}& \ghost{\mathcal{\shortstack{A\\D\\D\\E\\R}}} & \rstick{\ket{S\textsubscript{1,5} \rightarrow V\textsubscript{1,5}}} \qw\ &&&&&&&&\ghost{\mathcal{\shortstack{A\\D\\D\\E\\R}}}& \rstick{\ket{T\textsubscript{3,5}}}  \qw\\
&&&&&&&&\lstick{\ket{T\textsubscript{3,0}}}&  \qw\ & \qw\ &\qw\ & \qw\ & \qw\ &\qw\ & \qw\ &\qw\ & \qw\ &\ghost{\mathcal{\shortstack{A\\D\\D\\E\\R}}}\ & \rstick{\ket{P\textsubscript{6}}}  \qw\\
&&&&&&&&\lstick{\ket{T\textsubscript{3,1}}}&  \qw\ & \qw\ &\qw\ & \qw\ & \qw\ &\qw\ & \qw\ &\qw\ & \qw\ &\ghost{\mathcal{\shortstack{A\\D\\D\\E\\R}}}\ & \rstick{\ket{P\textsubscript{7}}}  \qw\\
&&&&&&&&\lstick{\ket{T\textsubscript{3,2}}}&  \qw\ & \qw\ &\qw\ & \qw\ & \qw\ &\qw\ & \qw\ &\qw\ & \qw\ &\ghost{\mathcal{\shortstack{A\\D\\D\\E\\R}}}\ & \rstick{\ket{P\textsubscript{8}}}  \qw\\
&&&&&&&&\lstick{\ket{T\textsubscript{3,3}}}&  \qw\ & \qw\ &\qw\ & \qw\ & \qw\ &\qw\ & \qw\ &\qw\ & \qw\ &\ghost{\mathcal{\shortstack{A\\D\\D\\E\\R}}}\ & \rstick{\ket{P\textsubscript{9}}}  \qw\\
&&&&&&&&\lstick{\ket{T\textsubscript{3,4}}}&  \qw\ & \qw\ &\qw\ & \qw\ & \qw\ &\qw\ & \qw\ &\qw\ & \qw\ &\ghost{\mathcal{\shortstack{A\\D\\D\\E\\R}}}\ & \rstick{\ket{P\textsubscript{10}}}  \qw\\
&&&&&&&&\lstick{\ket{T\textsubscript{3,5}}}&  \qw\ & \qw\ &\qw\ & \qw\ & \qw\ &\qw\ & \qw\ &\qw\ & \qw\ &\ghost{\mathcal{\shortstack{A\\D\\D\\E\\R}}}\ & \rstick{\ket{P\textsubscript{11}}}  \qw\\
}
\caption{Step 6: The arranged partial products derived from logical-ANDs are input to $2n-3$ bit adder \cite{C.Gidney.et.al-2018} is shown for 6-bit number $a$. $|T\rangle$ values are the partial products and input bits $a$ from Step 2. The sum qubit values $|S\rangle$ are the outputs from quantum adder. $|P\rangle$ values are the squared qubits ($a^{2}$) of input $a$.}
\label{Fig 7}
\end{figure}
\begin{itemize}
\item Step 6: \\
Step 6 takes place when $n$ is even. In this step, $(n-2)/2$ additions are to be performed. For each addition, two bits of the final squared value of $a$ are generated.  Each adder takes as input partial products stored in register $|T\rangle$ and the remaining sum bits from the last round of computation which is stored in quantum register $|V\rangle$.  The following executes lines 100 through 112 of algorithm 1. \\
An example of this Step is shown in Figure 7 for the case of 8 bit operands.\newline
\\
For i=1:(n-2)/2\newline
\parindent=15.0pt 
\indent For j=2n-2-2i
\begin{itemize}
\item Sub-Step 1: 
Apply $|V\textsubscript{i-1,0}\rangle$ to $|V\textsubscript{i-1,j-1}\rangle$ sum bits and $|T\textsubscript{i+1,0}\rangle$ to $|T\textsubscript{i+1,j-1}\rangle$ zero-padded partial products to $2n-2-2i$ bit  adder. On each addition operation  $|P\textsubscript{2i+2}\rangle$ and  $|P\textsubscript{2i+3}\rangle$ bits of the squared value $a^{2}$ are generated from the first two sum bits of the $2n-2-2i+1$ sum bits. The generated carry bit is redundant. \\
\item Sub-Step 2:
Place the remaining $2n-4-2i$ sum bits to quantum register  $|V\textsubscript{i,0}\rangle$ to  $|V\textsubscript{i,2n-5-2i}\rangle$. Repeat Sub-step 1.\\
\noindent If i=(n-2)/2 then $|P\textsubscript{2i+2}\rangle$ to $|P\textsubscript{2i+n+1}\rangle$ bits of the squared value  $|a\textsuperscript{2}\rangle$ are generated from $|V\textsubscript{i,0}\rangle$ to $|V\textsubscript{i,n-1}\rangle$.\\
\end{itemize}
\end{itemize}
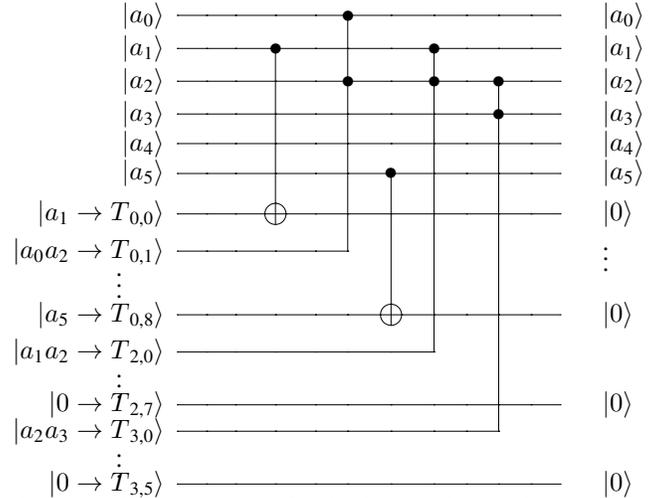
\begin{figure}[htpb]
\centering
\Qcircuit @C=1.1em @R=1em{
&&&&&&&&\lstick{\ket{a\textsubscript{0}}}& \qw\ & \qw\ & \qw\ & \qw\ & \ctrl{2}\ & \qw\ & \qw\ & \qw\ & \qw\ & \qw\ & \qw\ & \rstick{\ket{a\textsubscript{0}}} \\
&&&&&&&&\lstick{\ket{a\textsubscript{1}}}& \qw\ & \qw\ & \ctrl{5}\ &\qw\ & \qw\ & \qw\ & \ctrl{1}\ & \qw\ & \qw\ & \qw\ & \qw\ & \rstick{\ket{a\textsubscript{1}}} \\
&&&&&&&&\lstick{\ket{a\textsubscript{2}}}& \qw\ & \qw\ & \qw\ &\qw\ & \ctrl{5}\ & \qw\ & \ctrl{8}\ & \qw\ & \ctrl{1}\ & \qw\ & \qw\ & \rstick{\ket{a\textsubscript{2}}}\\
&&&&&&&&\lstick{\ket{a\textsubscript{3}}}& \qw\ & \qw\ & \qw\ &\qw\ & \qw\ & \qw\ & \qw\ & \qw\ &  \ctrl{10}\ & \qw\ & \qw\ & \rstick{\ket{a\textsubscript{3}}} \\
&&&&&&&&\lstick{\ket{a\textsubscript{4}}}& \qw\ & \qw\ & \qw\ &\qw\ & \qw\ & \qw\ & \qw\ & \qw\ & \qw\ & \qw\ & \qw\  & \rstick{\ket{a\textsubscript{4}}} \\
&&&&&&&&\lstick{\ket{a\textsubscript{5}}}& \qw\ & \qw\ & \qw\ &\qw\ & \qw\ & \ctrl{4}\ & \qw\ & \qw\ & \qw\ & \qw\ & \qw\ & \rstick{\ket{a\textsubscript{5}}} \\
&&&&&&&&\lstick{\ket{a\textsubscript{1}\rightarrow T\textsubscript{0,0}}}&\qw\ & \qw\ & \targ\ & \qw\ & \qw\ & \qw\ & \qw\ & \qw\ & \qw\ & \qw\ &\qw\ & \rstick{\ket{0}} \\
&&&&&&&&\lstick{\ket{a\textsubscript{0}a\textsubscript{2}\rightarrow T\textsubscript{0,1}}}&\qw\ & \qw\ & \qw\ & \qw\ & \qw\ &&&&&&&\rstick{\vdots\ } \\
&&&&&&\vdots\\
&&&&&&&&\lstick{\ket{a\textsubscript{5}\rightarrow T\textsubscript{0,8}}} &\qw\ & \qw\ & \qw\ & \qw\ & \qw\ & \targ\ & \qw\ & \qw\ & \qw\ & \qw\ & \qw\ & \rstick{\ket{0}} \\
&&&&&&&&\lstick{\ket{a\textsubscript{1} a\textsubscript{2}\rightarrow T\textsubscript{2,0}}} &\qw\ & \qw\ & \qw\  & \qw\ & \qw\ &\qw\  &\qw\\
&&&&&&\vdots\\
&&&&&&&&\lstick{\ket{0 \rightarrow T\textsubscript{2,7}}} &\qw\ & \qw\ & \qw\ & \qw\ & \qw\ & \qw\ & \qw\ & \qw\ & \qw\ & \qw\ & \qw\  &  \rstick{\ket{0}}\\
&&&&&&&&\lstick{\ket{a\textsubscript{2} a\textsubscript{3}\rightarrow T\textsubscript{3,0}}} & \qw\ & \qw\ & \qw\ & \qw\ & \qw\ & \qw\ & \qw\ & \qw\  & \qw\\
&&&&&&\vdots\\
&&&&&&&&\lstick{\ket{0\rightarrow T\textsubscript{3,5}}}&\qw\ & \qw\ & \qw\ & \qw\ & \qw\ & \qw\ & \qw\ & \qw\ & \qw\ & \qw\ & \qw\ & \rstick{\ket{0}}
}
\caption{Step 7 and 8 : The restoring of copied input qubits to their original ancillae value and the uncomputation of the partial products}
\label{Fig 7}
\end{figure}

\begin{itemize}
\item Step 7:\\
In this step the qubits of the copied values of input $a$ produced in step 1 are restored to their original ancillae values.\\
\parindent=15.0pt 
\indent For i=1:2:2n-3 \\
Apply input values $|a\textsubscript{(i+1)/2}\rangle$ and the quantum register value $|T\textsubscript{0,i-1}\rangle$ to CNOT-gate (line 115 of algorithm 2). At the end of computation, copied bits of input $a$ located in $|T\textsubscript{0,i-1}\rangle$ position will restored to initial value 0.\\
\end{itemize}

\begin{itemize}
\item Step 8:\\
In this step the original input bits of $a$ and the partial products produced in step 1 are restored to their original values of input $a$ (lines 1 to 40 of algorithm 2).
\\
If the number of input bits is greater than 4, then \\
For i=3:2n-3\\
\parindent=15.0pt
\begin{itemize}
\item Sub-step 1\\
\indent If $i\leq n-1$ and $i$ is odd then apply input $|a\textsubscript{1}\rangle$, $|a\textsubscript{i-1}\rangle$ and partial product $|T\textsubscript{2,i-3}\rangle$ to an uncomputation gate (line 7 of algorithm 2).
\begin{itemize}
\item Sub-sub-step 1\\
If i$>$3 then\\ 
\parindent=15.0pt
\indent For $i\textsubscript{1}=(i-1)/2$, apply input $|a\textsubscript{i\textsubscript{1}+1}\rangle$, $|a\textsubscript{i-i\textsubscript{1}-1}\rangle$ and partial product $|T\textsubscript{2+i\textsubscript{1},i-3-2*i\textsubscript{1}}\rangle$ to an uncomputation gate (line 10 of algorithm 2).
\end{itemize}
\item Sub-step 2\\
\indent If $i\leq n-1$ and $i$ is even then apply input $|a\textsubscript{0}\rangle$, $|a\textsubscript{i}\rangle$ and partial product $|T\textsubscript{0,i-1}\rangle$ to an uncomputation gate (line 15 of algorithm 2).
\begin{itemize}
\item Sub-sub-step 1	\\
If i$>$4 then\\ 
\parindent=15.0pt
\indent For $i\textsubscript{2}=(i/2)-2$, apply input $|a\textsubscript{i\textsubscript{2}+1}\rangle$, $|a\textsubscript{i\textsubscript{2}+1}\rangle$ and partial product $|T\textsubscript{2+i\textsubscript{2},i-1-2*i\textsubscript{2}}\rangle$ to an uncomputation gate (line 18 of algorithm 2).
\end{itemize}
\item Sub-step 3\\
If $i>n-1$ and $i\neq2n-3$, and $i$ is odd then,\\
\parindent=15.0pt
\indent For $i\textsubscript{3}=((2n-i-3)/2)$ apply input $|a\textsubscript{i-n+i\textsubscript{3}+1}\rangle$, $|a\textsubscript{n-i\textsubscript{3}-1}\rangle$ and partial product $|T\textsubscript{i\textsubscript{3}+1,i-1-2*i\textsubscript{3}}\rangle$ to an uncomputation gate (line 27 of algorithm 2).
\item Sub-step 4\\
If $i>n-1$ and $i$ is even then apply input  $|a\textsubscript{i-n+1}\rangle$, $|a\textsubscript{n-1}\rangle$ and partial product $|T\textsubscript{0,i-1}\rangle$ to an uncomputation gate (line 32 of algorithm 2).
\begin{itemize}
\item Sub-sub-step 1\\
If $i\neq2n-4$ and $i\neq2n-6$ then\\ 
\parindent=15.0pt
\indent For $i\textsubscript{4}=(2n-i-4)/2$ apply input$|a\textsubscript{i-n+i\textsubscript{4}+1]}\rangle$, $|a\textsubscript{n-i\textsubscript{4}-1]}\rangle$ and partial product $|T\textsubscript{i\textsubscript{4}+1,i-1-2*(i\textsubscript{4}-1)}\rangle$ to an uncomputation gate (line 35 of algorithm 2).\\
\end{itemize}
\end{itemize}
\end{itemize}

\section{Cost Analysis}

In this section the cost measures, namely T-count, T-depth, qubit cost, CNOT-count, CNOT-depth and KQ\textsubscript{T}, are calculated. The cost calculations are applicable for input bit $n>4$. Later the costs are compared in Table 1.

\subsection{T-gate Count}

For the proposed quantum squaring circuit, the T-count for even and odd number of input bits are separately illustrated for each step. The total T-count is calculated by adding the T-counts of each step of the proposed design methodology.\newline
\subsubsection{When $n$ is even}
\begin{itemize}
\item Step 1: T-count for the logical-AND:\newline
\newline
There are $\binom{n}{2}$ logical-AND operations. Each AND operation have a T-count of 4. The T-count for this step:\newline
$4n(n-1)/2 $.\newline

\item Step 2: T-count for ADDER:\newline
\newline
For even number of input bits $n$, we used a total of $n/2$ number of adders of varied sizes, where the sequence of the size of adder: \newline
\begin{equation}2n-3 + \sum_{i=0}^{\frac {n-4}{2}} 2n-4-2i\label{eq1}\end{equation}
Each $n$-bit adder has $n$ number of logical-ANDs. \newline
Hence, the total number of logical-ANDs present in $n/2$ number of adder for $n$-bit squaring circuit:\newline
\[2n-3 + \sum_{i=0}^{\frac {n-4}{2}} 2n-4-2i\]
\[=2n-3 + (3n^{2}-10n+8)/4\]
\begin{equation}=(3n^{2}-2n-4)/4\label{eq1}\end{equation}
Considering the total number of terms (adder) as $\frac{n-2}{2}$ for the series, the sum of the series is calculated. 
The T-count for this step : \begin{equation}4(3n^{2}-2n-4)/4\label{eq1} \end{equation}
 \newline
Therefore, for even number of input bits $n$, the total T-count required for the proposed QSC design:\newline
 \begin{equation} 4n(n-1)/2+ 4(3n^{2}-2n-4)/4= 5n^{2}-4n-4\label{eq1} \end{equation}
\end{itemize}
\subsubsection{When $n$ is odd}
\begin{itemize}
\item Step 1: T-count for the logical-AND:\newline
\newline
There are $\binom{n}{2}$ logical-AND operations. Each AND operation have a T-count of 4. The T-count for this step:\newline
$4n(n-1)/2 $.\newline

\item Step 2: T-count for ADDER:\newline
\newline
For odd number of input bits $n (n>4)$, we used a total of $(n-1)/2$ number of adders of varied sizes, where the sequence of the size of adder: \newline
\begin{equation}2n-3 + \sum_{i=0}^{\frac {n-5}{2}} 2n-4-2i\label{eq1}\end{equation}
Each $n$-bit adder has $n$ number of logical-ANDs. \newline
Hence, the total number of logical-ANDs present in $(n-1)/2$ number of adder for $n$-bit squaring circuit:\newline
\[2n-3 + \sum_{i=0}^{\frac {n-5}{2}} 2n-4-2i\] 
\[=2n-3+ (3n^{2}-12n+9)/4\] 
\begin{equation}=(3n^{2}-4n-3)/4\label{eq1}\end{equation}
Considering the total number of terms (adder) as $\frac {n-3}{2}$ for the series, the sum of the series is calculated. 
The T-count for this step : \begin{equation}4(3n^{2}-4n-3)/4\label{eq1} \end{equation}
 \newline
Therefore, for odd number of input bits $n (n>4)$, the total T-count required for the proposed QSC design:\newline
\begin{equation} 4n(n-1)/2+ 4(3n^{2}-4n-3)/4 = 5n^{2}-6n-3\label{eq1} \end{equation}
\end{itemize}

\subsection{T-depth Cost}
For the proposed quantum squaring circuit, the T-depth for even and odd number of input bits are separately illustrated shortly for each step. The total T-depth is calculated by adding the T-depth of each step of the proposed design methodology.\newline
\newline
\subsubsection{When $n$ is even}
\begin{itemize}
\item Step 1:  T-depth for the logical-AND:\newline
\newline
There are $\binom{n}{2}$ logical-AND operations. Each logical-AND operation have a T-depth of 2. The T-depth for this step: $2\binom{n}{2}$ =$n(n-1)$.\newline

\item Step 2: T-depth for the ADDER:\newline
\newline
From equation (2) we obtained, for even number of input bits $n(n>4)$ of our proposed QSC, total number of logical-ANDs present in $n/2$ number of adder: $(3n^{2}-2n-4)/4$.\newline
The T-depth for this step :  \begin{equation}2(3n^{2}-2n-4)/4\label{eq1} \end{equation}

Therefore, for even number of input bits $n(n>4)$, the total T-depth for the proposed QSC design: \newline
\begin{equation} n(n-1)+2(3n^{2}-2n-4)/4= (5n^{2}-4n-4)/2\label{eq1} \end{equation}
\end{itemize}
\subsubsection{When $n$ is odd}
\begin{itemize}
\item Step 1:  T-depth for the logical-AND:\newline
\newline
There are $\binom{n}{2}$ logical-AND operations. Each logical-AND operation have a T-depth of 2. The T-depth for this step: $2\binom{n}{2}$ =$n(n-1)$.\newline

\item Step 2: T-depth for the ADDER:\newline
\newline
From equation (5) we obtained, for odd number of input bits $n (n>4)$ of our proposed QSC, total number of logical-ANDs present in $(n-1)/2$ number of adder: $(3n^{2}-4n-3)/4$.\newline
The T-depth for this step:  \begin{equation}2(3n^{2}-4n-3)/4)\label{eq1} \end{equation}

Therefore, for odd number of input bits $n(n>4)$, the total T-depth for the proposed QSC design:\newline
\begin{equation}n(n-1)+ 2(3n^{2}-4n-3)/4= (5n^{2}-6n-3)/2\label{eq1} \end{equation}
\end{itemize}

\subsection{Qubit Cost}
The total qubit cost of the proposed design methodology is calculated by adding the total ancillae count to $n$-number of input bits. For the proposed quantum squaring circuit, the ancillae count is illustrated shortly for each step. The total ancillae is calculated by adding the ancillae of each step.\newline
\newline
\subsubsection {When $n$ is even}
\begin{itemize}
\item Step 1: Ancillae count for the logical-AND:\newline
\newline
There are $\binom{n}{2}$ logical-AND operations. Each AND operation have 1 ancillae. 1 ancillae is from 0 input. Total ancillae count for this step is   $\binom{n}{2}$ =\(n(n-1)/2+1\).\newline

\item Step 2: Ancillae count for ADDER: \newline
\newline
Each $n$-bit adder has $n$ logical-AND operations. Each AND operation have 1 ancillae.\newline
For even number of input bits $n(n>4)$, the total number of logical-ANDs in $n/2$ number of adder: $(3n^{2}-2n-4)/4$.\newline
Moreover, for the carry bit of the first adder an ancillae is used.
The ancillae count for this step : \begin{equation}(3n^{2}-2n-4)/4+1=(3n^{2}-2n)/4\label{eq1} \end{equation}
\item Step 3: Ancillae count for zero-padding:\newline
\newline
Zero-padding has been done twice for the proposed $n$-bit proposed QSC design:\newline
\begin{enumerate}
\item In between the partial products while arranging them in array:\newline
For even number of input bits $n(n>4)$, the location $|T\textsubscript{1,1}\rangle$, $|T\textsubscript{1,2n-5}\rangle$ and $|T\textsubscript{n/2,1}\rangle$ of the arranged partial product will have three zeroes. The remaining $|T\textsubscript{i,j}\rangle$ locations where $i=2,3,....(n-2)/2$ and $j=0,1,....(4n-2i+2)$ will have $3(n-4)/2$ number of zeros. The ancillae count for this step:\\
 \begin{equation}3+3(n-4)/2=(3n-6)/2\label{eq1} \end{equation}
\item To the left side of the partial products after arranging them in array:\newline
For even number of input bits $n$ the zero-padding sequence: \\
\[\sum_{i=1}^{\frac {n-2}{2}} 2i=(n^{2}-2n)/4\]\newline
The ancillae count for this step:  \begin{equation}(n^{2}-2n)/4\label{eq1} \end{equation}
The total ancillae count is:\\
$(n^{2}-n)/2+ (3n^{2}-2n)/4 + (n^{2}-2n)/4+(3n-6)/2+1$\\
\begin{equation}= (3n^{2} - 4)/2\label{eq1} \end{equation}
Therefore, for even number of input bits $n$, the total qubit cost required for the proposed QSC design:  \textit{n (no. of qubits) + Ancillae count}\\
\begin{equation} =(3n^{2} + 2n - 4)/2\label{eq1} \end{equation}
\end{enumerate} 
\end{itemize}
\subsubsection {When $n$ is odd}
\begin{itemize}
\item Step 1: Ancillae count for the logical-AND:\newline
\newline
There are $\binom{n}{2}$ logical-AND operations. Each AND operation have 1 ancillae. 1 ancillae is from 0 input. Total ancillae count for this step is   $\binom{n}{2}$ =\(n(n-1)/2+1\).\newline
\newline
\item Step 2: Ancillae count for ADDER: \newline
\newline
Each $n$-bit adder has $n$ logical-AND operations. Each AND operation have 1 ancillae. 1 ancillae is from 0 input.\newline
For odd number of input qubits $n$, the total number of logical-ANDs in $(n-1)/2$ number of adder: $(3n^{2}-4n-3)/4$.\newline
Moreover, for the carry bit of the first adder an ancillae is used.
The ancillae count for this step: \begin{equation}(3n^{2}-4n-3)/4 + 1= (3n^{2}-4n+1)/4\label{eq1} \end{equation}
\item Step 3: Ancillae count for zero-padding:\newline
\newline
Zero-padding has been done twice for the proposed $n$-bit proposed QSC design:\newline
\begin{enumerate}
\item In between the partial products while arranging them in array:\newline
For odd number of input bits $n(n>4)$, the locations $|T\textsubscript{1,1}\rangle$, $|T\textsubscript{1,2n-5}\rangle$, $|T\textsubscript{(n-1)/2,1}\rangle$ and $|T\textsubscript{(n-1)/2,3}\rangle$ of the quantum register $|T\rangle$ will have four zeroes. The remaining $|T\textsubscript{i,j}\rangle$ locations where $i=2,3,....(n-3)/2$ and $j=0,1,....(2n-4i+2)$ will have $3(n-5)/2$ number of zeros. 
The ancillae count for this step:\\
\begin{equation}4+3(n-5)/2=(3n-7)/2\label{eq1} \end{equation}
\item To the left side of the partial products after arranging them in array:\newline
For odd number of input bits $n(n>4)$ the zero-padding sequence: \\
\[\sum_{i=1}^{\frac {n-3}{2}} 2i=(n^{2}-4n+3)/4\]\newline
The ancillae count for this step: \begin{equation}(n^{2}-4n+3)/4\label{eq1} \end{equation}
The total ancillae count is:\\
$(n^{2}-n)/2+ (3n^{2}-4n+1)/4 + (n^{2}-4n+3)/4 +(3n-7)/2+1$\\
\begin{equation}=(3n^{2} - 2n -3)/2\label{eq1} \end{equation}
Therefore, for odd number of input bits $n$, the total qubit cost for the proposed QSC design:  \textit{n (no. of qubits) + Ancillae count}\\
\begin{equation}= (3n^{2} - 3)/2\label{eq1} \end{equation}
\end{enumerate} 
\end{itemize}

\subsection{CNOT-gate Count}
\subsubsection {When $n$ is even}
\begin{itemize}
\item Step 1: CNOT-count for the logical-AND:\newline
\newline
There are $\binom{n}{2}$ logical-ANDs. Each AND has 6 CNOT-gates. There are $n-1$ CNOT-gates to copy the input bits and then restore to 0.\newline
The CNOT-count for this step: $2(n-1)+6 \binom{n}{2}=3n^{2}-n-2$. \newline
 
\item Step 2: CNOT-count for ADDER:\newline
\newline 
For even number of input bits $n (n>4)$, we used a total of $n/2$ number of adders of varied sizes, where the sequence of the size of adder: \newline
 \begin{equation}2n-3 + \sum_{i=0}^{\frac {n-4}{2}} 2n-4-2i \label{eq1} \end{equation}
For n-bit of  adder there are $12n-9$ CNOT-gates.\newline
For $2n-3$ bits of adder there are $24n-45$ CNOT-gates.\newline
For $2n-4$ bits of adder there are $24n-57$ CNOT-gates.\newline
Hence, the total number of CNOT-gates present in $n/2$ number of adder for $n$-bit squaring circuit:\newline
\[ 24n-45 + \sum_{i=0}^{\frac {n-4}{2}} 24n-57-12i \] 
\[=24n-45+ (18n^{2}-69n+66)/2 \] 
 \begin{equation}= (18n^{2}-21n-24)/2 \label{eq1} \end{equation} 
Considering the total number of terms (adder) as $\frac{n-2}{2}$ for the series, the sum of the series is calculated. 
 \newline
Therefore, for even number of input bits $n (n>4)$, the total number of CNOT-gates for the proposed QSC design: \newline
 \begin{equation}3n^{2}-2n + (18n^{2}-21n-24)/2 = (24n^{2}-23n-28)/2 \label{eq1} \end{equation}
\end{itemize}

\subsubsection {When $n$ is odd}
\begin{itemize}
\item Step 1: CNOT-count for the logical-AND:\newline
There are $\binom{n}{2}$ logical-ANDs. Each AND has 6 CNOT-gates. There are $n-1$ CNOT-gates to copy the input bits and restore to 0.\newline
 The CNOT-count for this step: $2(n-1)+6 \binom{n}{2}=3n^{2}-n-2$. \newline
 \newline
\item Step 2: CNOT-count for  ADDER:\newline
For odd number of input bits $n$, we used a total of $(n-1)/2$ number of adders of varied sizes, where the sequence of the size of adder: \newline
 \begin{equation}2n-3 + \sum_{i=0}^{\frac {n-4}{2}} 2n-4-2i \label{eq1} \end{equation}
For n-bit of  adder there are $12n-9$ CNOT-gates.\newline
For $2n-3$ bits of adder there are $24n-45$ CNOT-gates.\newline
For $2n-4$ bits of adder there are $24n-57$ CNOT-gates.\newline
Hence, the total number of CNOT-gates present in $(n-1)/2$ number of adder for $n$-bit squaring circuit:\newline
\[ 24n-45 + \sum_{i=0}^{\frac {n-5}{2}} 24n-57-12i \] 
\[=24n-45+ (18n^{2}-81n+81)/2 \] 
 \begin{equation}= (18n^{2}-33n-9)/2 \label{eq1} \end{equation}
Considering the total number of terms (adder) as $\frac{n-3}{2}$ for the series, the sum of the series is calculated. 
 \newline
Therefore, for odd number of input bits $n$, the total number of CNOT-gates for the proposed QSC design: \newline
 \begin{equation}3n^{2}-n-2 + (18n^{2}-33n-9)/2 = (24n^{2}-35n-13)/2 \label{eq1} \end{equation}
\end{itemize}

\subsection{CNOT-depth Cost}
\subsubsection {When $n$ is even}
\begin{itemize}
\item Step 1: CNOT-depth for the logical-AND:\newline
There are $\binom{n}{2}$ logical-AND. Each AND has a CNOT-depth of 4. There are  $n-1$ CNOT-gates to copy the input bits and then restore to 0. The CNOT-depth for this step: $2(n-1)+4\binom{n}{2}$=$2n^{2}-2$.\newline

\item Step 2: CNOT-depth for  ADDER:\newline
For even number of input bits $n (n>4)$, we used a total of $n/2$ number of adders of varied sizes, where the sequence of the size of adder: \newline
 \begin{equation}2n-3 + \sum_{i=0}^{\frac {n-4}{2}} 2n-4-2i\label{eq1} \end{equation}
CNOT-depth of $n$-bit of adder is $8n-6$.\newline
CNOT-depth of $2n-3$ bits of adder is $16n-30$.\newline
CNOT-depth of $2n-4$ bits of adder is $16n-38$.\newline
\newline
Hence, the total CNOT-depth of $n/2$ number of adder for $n$-bit squaring circuit:\newline
\[ 16n-30 + \sum_{i=0}^{\frac {n-4}{2}} 16n-38-16i \] 
\[=16n-30+ 6n^{2}-23n+22 \] 
 \begin{equation}= 6n^{2}-7n-8\label{eq1} \end{equation}
Considering the total number of terms (adder) as $\frac{n-2}{2}$ for the series, the sum of the series is calculated. 
 \newline
Therefore, for even number of input bits $n(n>4)$, the total CNOT-depth for the proposed QSC design:\\
 \begin{equation} 2n^{2}-2 +6n^{2}-7n-8= 8n^{2}-7n-10 \label{eq1} \end{equation}
\end{itemize}

\subsubsection {When $n$ is odd}
\begin{itemize}
\item Step 1: CNOT-depth for the logical-AND:\newline
\newline
There are $\binom{n}{2}$ logical-AND. Each AND has a CNOT-depth of 4. There are $n-1$ CNOT-gates to copy the input bits and then restore to 0. The CNOT-depth for this step: $2(n-1)+4\binom{n}{2}$=$2n^{2}-2$.\newline

\item Step 2: CNOT-depth for  ADDER:\newline
\newline
For odd number of input bits $n$, we used a total of $(n-1)/2$ number of adders of varied sizes, where the sequence of the size of adder: \newline
 \begin{equation}2n-3 + \sum_{i=0}^{\frac {n-4}{2}} 2n-4-2i\label{eq1} \end{equation}
CNOT-depth of $n$-bit of adder is $8n-6$.\newline
CNOT-depth of $2n-3$ bits of adder is $16n-30$.\newline
CNOT-depth of $2n-4$ bits of adder is $16n-38$.\newline
\newline
Hence, the total CNOT-depth of $(n-1)/2$ number of adder for $n$-bit squaring circuit:\newline
\[ 16n-30 + \sum_{i=0}^{\frac {n-5}{2}} 16n-38-16i \] 
\[=16n-30+ 6n^{2}-27n+27 \] 
 \begin{equation}= 6n^{2}-11n-3\label{eq1} \end{equation}
Considering the total number of terms (adder) as $\frac{n-3}{2}$ for the series, the sum of the series is calculated. 
 \newline
Therefore, for odd number of input bits $n (n>4)$, the total CNOT-depth for the proposed QSC design: \newline
 \begin{equation}2n^{2}-2 + 6n^{2}-11n-3=8n^{2}-11n-5\label{eq1} \end{equation}
\end{itemize}

\subsection{KQ\textsubscript{T} Cost}
The KQ\textsubscript{T} cost is the product of qubit cost and T-depth. The KQ\textsubscript{T} for the proposed QSC design:\newline
\newline
\subsubsection {When $n$ is even}
For even number of input bits $n$ KQ\textsubscript{T} for the proposed QSC design:\\
\\ 
$(3n^{2} + 2n - 4)/2 \cdot (5n^{2}-4n-4)/2$ \\
\begin{equation}=(15n^{4}- 2n^{3}-40n^{2}+8n+16)/4\label{eq1}\end{equation}
\newline
\subsubsection {When $n$ is odd}
For odd number of input bits $n$ KQ\textsubscript{T} for the proposed QSC design:\\
\\
$(3n^{2}- 3)/2 \cdot (5n^{2}-6n-3)/2$
\begin{equation}=(15n^{4}-18n^{3}-24n^{2}+18n+9)/4\end{equation}

\begin{table*}[htpb]
\centering
\caption{Cost Comparison of Quantum Squaring Circuits}
\resizebox{\textwidth}{!}
{
\begin{tabular}{*{4}{c}}
\\ \hline
\textbf{Cost Measures} & \textbf{\textit{Thapliyal et al.}}& \textbf{\textit{(OSU) Nagamani et al.}}& \textbf{\textit{Proposed QSC}} \\
\\ \hline
T-count & $15n^{2}-17n+2$ & $22n^{2} –24n-12$ & n-even: $5n^{2}– 4n-4$\\ &&& n-odd: $5n^{2}–6n–3$ \\
\\ 
Qubit Cost & $n^{2}+2n+1$ & $(n^{2}+5n+4)/2$ & n-even: $(3n^{2}+2n-4)/2$\\ &&&  n-odd:  $(3n^{2}-3)/2$\\
\\
T-depth & $5n^{2}-3n-2$ & $8n^{2}–6n-8$ & n-even: $(5n^{2}-4n-4)/2$\\&&&  n-odd: $(5n^{2} – 6n-3)/2$\\
\\
CNOT-count & $17n^{2}-23n+8$ & $24n^{2}–52n-6$ & n-even: $(24n^{2}-23n-28)/2 $\\ &&&  n-odd: $ (24n^{2}-35n-13)/2 $\\
\\
CNOT-depth & $14n^{2}-14n+2$ & $21n^{2}–21n-12$ & n-even: $8n^{2}-7n-10$\\ &&&  n-odd: $8n^{2}-11n-5$\\
\\
KQ\textsubscript{T} & $5n^{4}+7n^{3}-3n^{2}-7n-2$ & $4n^{4}+17n^{3}-3n^{2}-32n-16$ & n-even: $ (15n^{4}- 2n^{3}-40n^{2}+8n+16)/4$\\ &&&  n-odd: $(15n^{4}-18n^{3}-24n^{2}+18n+9)/4$\\
\hline
\end{tabular}
\label{tab1}
}
\end{table*}

\section{Cost Comparison}
In this section we compare the cost of our proposed QSC design with other existing works. The works presented in \cite{banerjee2014squaring}, \cite{banerjee2016squaring}, and \cite{jayashree2017efficient} are not considered because these works have higher gate and depth costs than \cite{Thapliyal.et.al-2014} and \cite{Nagamani.et.al-2013}. Our proposed QSC is composed of logical-AND gates and the adder presented in \cite{C.Gidney.et.al-2018}.  The adder in \cite{C.Gidney.et.al-2018} is composed of logical-AND gates and CNOT-gates. These logical-AND gates can be decomposed into Clifford + T-gates and CNOT-gates as shown in Fig. 1. The cost calculation of the existing works is done in terms of quantum gates and ancillae. \newline
Since our circuit uses the logical-AND gate that can be decomposed into Clifford + T-gates, we decomposed the Toffoli, Peres Gate and Double Peres Gate used in \cite{Thapliyal.et.al-2014} and \cite{Nagamani.et.al-2013} into Clifford +T-gates as for a better comparison. For the existing designs, the implementations of the Toffoli gate requires seven T-gates, the Peres Gate requires seven T-gates and the Double Peres Gate requires eight T-gates \cite{Thapliyal.et.al-2014}\cite{Nagamani.et.al-2013} whereas in our design the Clifford + T implementations of the logical-AND gate requires four T-gates and the n-bit adder presented in \cite{C.Gidney.et.al-2018} requires $4(n-1)$ T-gates.\newline
\newline
The design presented in \cite{Thapliyal.et.al-2014} is made garbage free by Nagamani et al. and presented as the first squaring design titled Garbage Free Squaring Unit \textit{(GFSU)} in \cite{Nagamani.et.al-2013}. Nagamani et al. presented another garbage optimized quantum squaring unit in \cite{Nagamani.et.al-2013} titiled \textit{Optimized Squaring Unit (OSU)}. For precise comparison we made this \textit{Optimized Squaring Unit (OSU)} presented in \cite{Nagamani.et.al-2013} garbageless by applying the Bennett’s garbage removal scheme \cite{bennett1973logical}. Consequently, the garbageless forms of the \textit{Optimized Squaring Unit (OSU)} by Nagamani et al. \cite{Nagamani.et.al-2013} requires $2 . n + 1$ additional qubits and sees an increase in the T-count by a factor of $2$ and CNOT-count by a factor of $2$, plus an additional $2n$ CNOT-gates for copying the output i.e. $(2+2n)$.\newline
\newline
We calculate the cost by taking the asymptotic limit of the proposed QSC and compare it with \cite{Thapliyal.et.al-2014} and \cite{Nagamani.et.al-2013}. For example, in order to compare the T-gate cost of proposed QSC with Thapliyal et. al. \cite{Thapliyal.et.al-2014}:
The T-count of proposed QSC is $5n^{2}-4n-4$. The T-count of Thapliyal et. al. \cite{Thapliyal.et.al-2014} is $15n^{2}-17n+2$. \newline
\newline
The asymptotic limit of proposed QSC:  $\underset{n\to \infty}{\lim}(5n^{2}-4n-4)$ and Thapliyal et. al. is $\underset{n\to \infty}{\lim}(15n^{2}-17n+2)$. \newline
\newline
The ratio of the cost: $\frac{\underset{n\to \infty}{\lim}(5n^{2}-4n-4)}{\underset{n\to \infty}{\lim}(15n^{2}-17n+2)} \approx \frac{5}{15}=\frac{1}{3}$. \newline
\newline
Therefore, the asymptotic reduction in T-count of proposed QSC design with respect to Thapliyal et. al. \cite{Thapliyal.et.al-2014}: $\frac{3-1}{3}=66.67\%$ 

Again, in order to compare the T-gate cost of proposed QSC with Nagamani et. al. \cite{Nagamani.et.al-2013}:
The T-count of proposed QSC is $5n^{2}-4n-4$. The T-count of Nagamani et. al. \cite{Nagamani.et.al-2013} is $22n^{2}-24n-12$. \newline
\newline
The asymptotic limit of proposed QSC:  $\underset{n\to \infty}{\lim}(5n^{2}-4n-4)$ and Nagamani et. al. is $\underset{n\to \infty}{\lim}(22n^{2}-24n-12)$. \newline
\newline
The ratio of the cost: $\frac{\underset{n\to \infty}{\lim}(5n^{2}-4n-4)}{\underset{n\to \infty}{\lim}(22n^{2}-24n-12)} \approx \frac{5}{22}$. \newline
\newline
Therefore, the asymptotic reduction in T-count of proposed QSC design with respect to Nagamani et. al. \cite{Nagamani.et.al-2013}: $\frac{22-5}{22}=77.27\%$

\subsection{Cost Comparison in Terms of T-count}
Table 1 illustrates that the order of growth of the T-count for the proposed QSC is quadratic. Thus, the T-count is $O(n^{2})$. Table 1 also illustrates that the T-count for the designs presented by Thapliyal et al. \cite{Thapliyal.et.al-2014}, and Nagamani et al. \cite{Nagamani.et.al-2013} are also $O(n^{2})$. By taking the ratio of the asymptotic limit of the T count polynomials for our proposed QSC and the designs presented by Thapliyal et al. \cite{Thapliyal.et.al-2014} and Nagamani et al. \cite{Nagamani.et.al-2013} we found that our proposed QSC shows 66.67\% reduction in T-count than the design Thapliyal et al. \cite{Thapliyal.et.al-2014} and 77.27\% reduction in T-count than the design by Nagamani et al. \cite{Nagamani.et.al-2013}. 

\subsection{Cost Comparison in Terms of T-depth}
Table 1 shows an inverse relation between T-depth and qubit cost. Minimizing T-depth will result in an increase in the qubit cost and vice-versa. 
Table 1 illustrates that the order of growth of the T-depth for the proposed QSC is quadratic. Thus, the T-depth is $O(n^{2})$. Table 1 also illustrates that the T-depth for the designs presented by Thapliyal et al. \cite{Thapliyal.et.al-2014} and Nagamani et al. \cite{Nagamani.et.al-2013} are also $O(n^{2})$. By taking the ratio of the asymptotic limit of the T-depth polynomials for our proposed QSC and the designs presented by Thapliyal et al. \cite{Thapliyal.et.al-2014} and Nagamani et al. \cite{Nagamani.et.al-2013} we found that our proposed QSC shows 50\% reduction in T-depth than the design Thapliyal et al. \cite{Thapliyal.et.al-2014} and 68.75\% reduction in T-depth than the design by Nagamani et al. \cite{Nagamani.et.al-2013}. Our proposed design achieves a qubit cost of order$O(n^{2})$, which is the same as the prior art.  Thus, we reduce the overall T-depth cost and maintained a qubit cost of the same order as prior art.

\subsection{Cost Comparison in Terms of CNOT-count}
Table 1 illustrates that the order of growth of the CNOT-count for the proposed QSC is quadratic. Thus, the CNOT-count is $O(n^{2})$. Table 1 also illustrates that the CNOT-count for the designs presented by Thapliyal et al. \cite{Thapliyal.et.al-2014}, and Nagamani et al. \cite{Nagamani.et.al-2013} are also $O(n^{2})$. By taking the ratio of the asymptotic limit of the CNOT-count polynomials for our  proposed QSC and the designs presented by Thapliyal et al. \cite{Thapliyal.et.al-2014} and Nagamani et al. \cite{Nagamani.et.al-2013} we found that our proposed QSC shows 29.41\% reduction in CNOT-count than the design Thapliyal et al. \cite{Thapliyal.et.al-2014} and 50\% reduction in CNOT-count than the design by Nagamani et al. \cite{Nagamani.et.al-2013}.

\subsection{Cost Comparison in Terms of CNOT-depth}
Table 1 illustrates that the order of growth of the CNOT-depth for the proposed QSC is quadratic. Thus, the CNOT-depth is $O(n^{2})$. Table 1 also illustrates that the CNOT-depth for the designs presented by Thapliyal et al. [1] and Nagamani et al. [2] are also $O(n^{2})$. By taking the ratio of the asymptotic limit of the CNOT-depth polynomials for our proposed QSC and the designs presented by Thapliyal et al. \cite{Thapliyal.et.al-2014} and Nagamani et al. \cite{Nagamani.et.al-2013} we found that our proposed QSC design 1 shows 42.86\% reduction in CNOT-depth than the design Thapliyal et al. \cite{Thapliyal.et.al-2014} and 61.90\% reduction in CNOT-depth than the design by Nagamani et al. \cite{Nagamani.et.al-2013}. We reduce the overall CNOT-depth cost than all the exisiting designs and maintained a qubit cost of the same order as prior art.

\subsection{Cost Comparison in Terms of KQ\textsubscript{T}}
Table 1 illustrates that the order of growth of the KQ\textsubscript{T} for the proposed QSC is a polynomial of degree four. Thus, the KQ\textsubscript{T} is $O(n^{4})$. Table 1 also illustrates that the KQ\textsubscript{T} for the designs presented by Thapliyal [1], and Nagamani et al. [2] are also $O(n^{4})$. By taking the ratio of the asymptotic limit of the KQ\textsubscript{T} polynomials for our proposed QSC and the designs presented by Thapliyal et al. \cite{Thapliyal.et.al-2014} and Nagamani et al. \cite{Nagamani.et.al-2013} we found that our proposed QSC shows 25\% reduction in KQ\textsubscript{T} than the design Thapliyal et al. \cite{Thapliyal.et.al-2014} and 6.25\% reduction in KQ\textsubscript{T} than the design by Nagamani et al. \cite{Nagamani.et.al-2013}. The work achieves a lower KQ\textsubscript{T} measure against all the existing works despite the constant factor increase in qubit cost compared to the prior art.

\section{Conclusion}
In this work, we present a novel design for quantum squaring.  The proposed design has no garbage output. The proposed work also enjoys  lower T-count, T-depth, CNOT-count, CNOT-depth and  KQ\textsubscript{T}  costs compared to the current state of the art.  The proposed squaring circuit has been verified for correctness.    We conclude that our proposed QSC design could be an ideal building block for quantum implementations of scientific computing or encryption algorithms where the T-gate and CNOT-gate costs must be kept to minimum.

\bibliographystyle{IEEEtran}
\bibliography{ref2}


 





\end{document}